\documentclass{cernyrep} 
\pdfoutput=1
\usepackage{texnames}
\usepackage[T1]{fontenc}
\begin{document}

\title{Quantum Field Theory and the Electroweak Standard Model\footnote{Lectures 
given at the European School of High-Energy Physics (ESHEP), 
September 2017, Evora, Portugal}
}
 
\author{A.B. Arbuzov}

\institute{BLTP JINR, Dubna, Russia}

\maketitle % this produces the title block

\begin{abstract}
Lecture notes with a brief introduction to Quantum field theory and 
the Standard Model are presented. The lectures were given at the
2017 European School of High-Energy Physics. 
The main features, the present status, and problems of the Standard Model 
are discussed. 
\end{abstract}
 
\section{Introduction
\label{sec:Intro}}

The lecture course consists of four main parts. In the Introduction, we will discuss what
is the Standard Model (SM)~\cite{Glashow:1961tr,Weinberg:1967tq,Salam:1968rm}, 
its particle content, and the main principles of its construction. 
The second Section contains brief notes on Quantum Field Theory (QFT), 
where we remind the main objects and rules required further for construction of the SM.
Sect.~\ref{sect:SM} describes some steps of the SM development. The Lagrangian of
the model is derived and discussed.
Phenomenology and high-precision tests of the model are overviewed in Sect.~\ref{sect:Pheno}.
The present status, problems, and prospects of the SM are summarized in Conclusions.  
Some simple exercises and questions are given for students in each Section.
These lectures give only an overview of the subject while for details one should look
in textbooks, \eg \cite{Bogoliubov_Shirkov,Ryder:1985wq,Weinberg_book,Peskin_Schroeder}, 
and modern scientific papers.

\subsection{What is the Standard Model?
\label{sec:Intro1}}

Let us start with the definition of the main subject of the lecture course.
It is the so-called \emph{Standard Model}. This name is quite widely accepted
and commonly used to define a certain theoretical model in high energy physics.
This model is suited to describe properties and interactions of \emph{elementary}
particles. One can say that at the present moment, the Standard Model is the most
successful physical model ever. In fact it describes with a high precision 
hundreds and hundreds independent observables. The model made also a lot of
predictions which have been verified later experimentally. Among other physical
models pretending to describe fundamental properties of Nature, the SM has the
highest predictive power. Moreover, the model is minimal: it is constructed 
using only fields, interactions, and parameters which are necessary for consistency
and/or observed experimentally. 
The minimality and in general the success of the model is provided to a great extent
by application of symmetry principles.

In spite of the nice theoretical features and successful experimental
verification of the SM, we hardly can believe that it is the true
fundamental theory of Nature. First of all, it is only one of an infinite
number of possible models within Quantum field theory. So it has
well defined grounds but its uniqueness is questionable. Second, we will
see that the SM and QFT itself do not seem to be the most adequate (mathematical)
language to describe Nature. 
One can also remind that gravity is not (yet) joined uniformly with
the SM interactions.

In any case, the SM is presently the main theoretical tool in high-energy physics.
Most likely this status will be preserved even if some new more fundamental physical 
model would be accepted by the community. In this case the SM will be treated as
an approximation (a low-energy limit) of that more general theory. But for practical
applications (in a certain energy domain) we will still use the SM.

\subsection{Particle content of the Standard Model
\label{sec:Intro2}}

Before construction of the SM, let us defined its content in
the sense of fields and particles. 

We would like to underline that the discovery of the Higgs
boson at LHC in 2012~\cite{Aad:2012tfa,Chatrchyan:2012xdj}
just finalized the list of SM particles from the experimental
point of view. Meanwhile the Higgs boson is one of the key
ingredients of the SM, so it was always in the list even so that
its mass was unknown.

\begin{figure}
\centering\includegraphics[width=.9\linewidth]{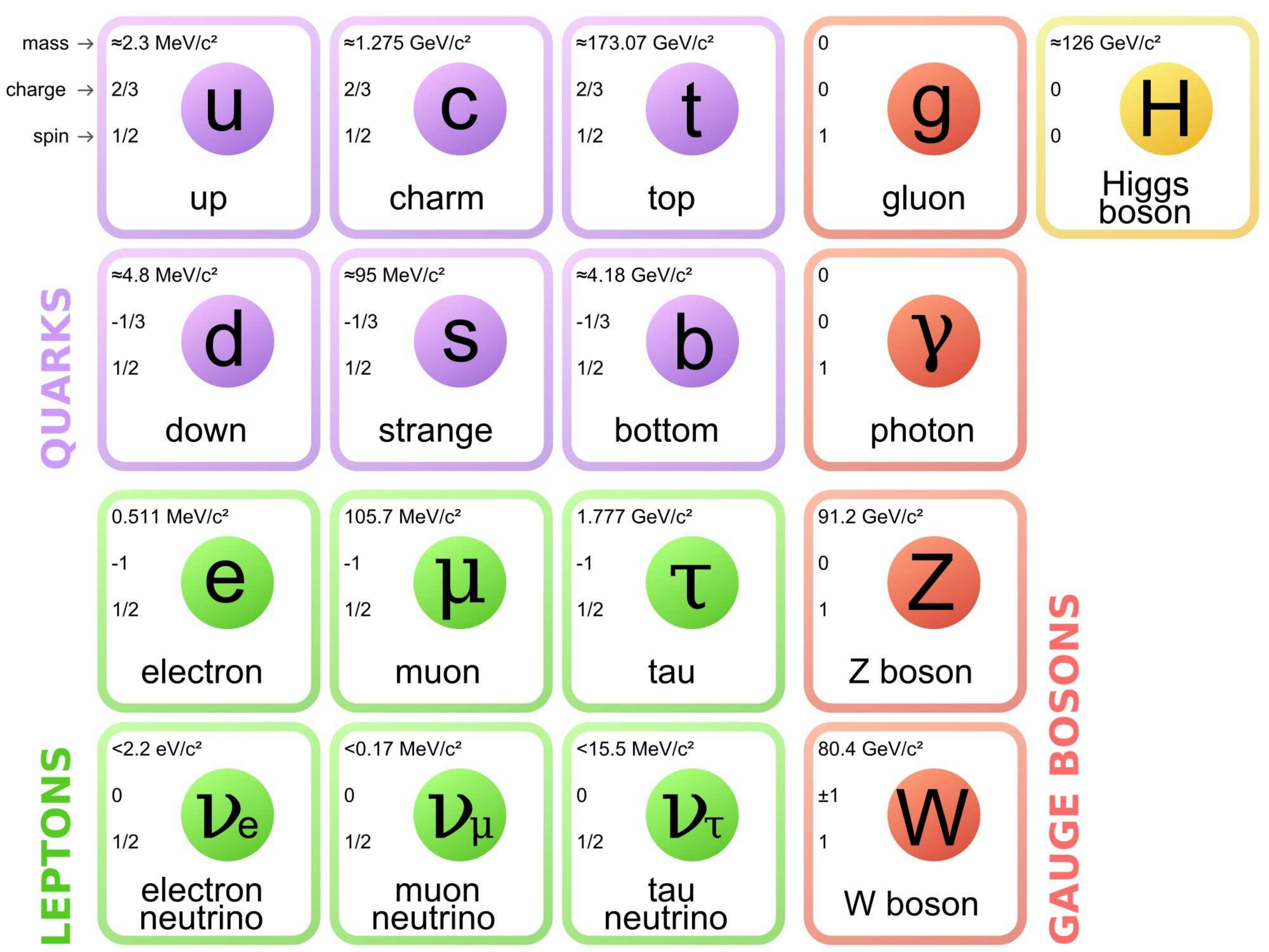}
\caption{Particle content of the Standard Model. 
Courtesy to Wikipedia: 'Standard Model of Elementary Particles' by MissMJ --- Own work by uploader, 
PBS NOVA, Fermilab, Office of Science, United States Department of Energy, Particle Data Group.}
\label{fig:SM_particles}
\end{figure}

The particle content of the SM is given on~\Fref{fig:SM_particles}. 
It consists of 12 fermions (spin = 1/2), 
4 vector gauge bosons (spin = 1), 
and one scalar Higgs boson (spin = 0). 
For each particle the chart contains information
about its mass, electric charge, and spin.
One can see that the data on neutrino masses is represented in the form
of upper limits, since they have not been yet measured.
Strictly speaking the information about neutrino masses should be treated
with care. According to the present knowledge, as discussed in the course
of lectures on Neutrino Physics, a neutrino particle of a given lepton flavor
\eg $\nu_\tau$, is not a mass eigenstate but a superposition of 
(at least) three states with different masses. 
    
Fermions are of two types: \emph{leptons} and \emph{quarks}. 
They are: \\
--- 3 charged leptons ($e$, $\mu$, $\tau$); \\
--- 3 neutrinos $\nu_e$, $\nu_\mu$, $\nu_\tau$  
(or $\nu_1$, $\nu_2$, $\nu_3$, see lectures on Neutrino physics); \\
--- 6 quarks of different \emph{flavors}, see lectures on Flavor Physics.

Each quark can have one of three \emph{colours}, see lectures on QCD.
Each fermion has 2 degrees of freedom \eg can have spin up or down,
or can be either \emph{left} or \emph{right}.
Each fermion particle in the SM has an \emph{anti-particle}, $f\neq \bar{f}$.     
The later statement is not yet verified for neutrinos, they might be Majorana
particles. 

Traditionally fermions are called \emph{matter fields}, contrary the the so-called
\emph{ force fields}, \ie intermediate vector bosons which mediate gauge interactions. 
Please keep in mind that this notion doesn't correspond to the common sense directly.
In fact most of fermions are unstable and do not form the 'ordinary matter' around
us, while \eg the mass of nuclear matter is provided to a large extent by gluons.
Moreover, looking at various Feynman diagrams we can see that fermions can serve as
intermediate particles in interaction processes.

In the SM we have the following boson fields: \\
--- 8 vector (spin=1) \emph{gluons}; \\
--- 4 vector (spin=1) \emph{electroweak bosons}: $\gamma$, $Z$, $W^+$, $W^-$; \\
--- 1 scalar (spin=0) \emph{Higgs boson}.

Gluons and photon are \emph{massless} and have 2 degrees of freedom (polarizations),
$Z$ and $W$ bosons are \emph{massive} and have 3 degrees of freedom (polarizations). 
By saying massless or massive we mean the absence or presence of the corresponding
terms in the Lagrangian of the SM. This is not always related to observables
in a straightforward way: \eg gluons are not observed as free asymptotic massless
states, and  masses of unstable $W$ and $Z$ bosons are defined indirectly from 
kinematics of their decay products.

Gluons and Electroweak (EW) bosons are \emph{gauge bosons}, their interactions with 
fermions are fixed by certain symmetries of the SM Lagrangian.
Note that electrically neutral bosons $(H$, $\gamma$, $Z$, and gluons) coincide with
their anti-particles \eg $\gamma\equiv\bar{\gamma}$. 
Each of 8 gluons carries one color and one anti-color.

Besides the particle content, we have to list the interactions which are described by 
the Standard Model. One of our final ultimate goals would be to answer the question 
\emph{ ``How many fundamental interactions are there in Nature?''}
But we should understand that it is only a dream, a primary motivation of our
studies. Being scientists we should be always unsure about the true answer to this
question. On the other hand, we can certainly say, how many different interactions
is there in a given model, for example in the SM. 
To answer this question we have to look at the complete Lagrangian of the model, 
see \eg book~\cite{Bardin:1999ak}.
For the SM it looks very long and cumbersome. The SM Lagrangian
contains kinetic terms for all listed above fields and dozens of terms
that describe interactions between them. So, before trying to count the 
number of interactions we should understand the structure and symmetries
of the Lagrangian.

\subsection{Principles of the Standard Model
\label{sec:Intro3}}
 
We are going to construct the SM Lagrangian. For this purpose, we have to
define first the guiding principles. That is important for optimization
of the procedure. The same principles might be used further in construction
of other models. 

First of all, we have to keep in mind that the SM is a model that is built 
within the local Quantum field theory. From the beginning this condition strongly
limits the types of terms that can appear in the Lagrangian because of the
Lorentz invariance, the Hermitian condition, the locality \etc
One can make a long list of various conditions. 
Here I list only the main principles which will be exploited in our way of
the SM construction:
\begin{itemize}
\item{the \emph{generalized correspondence} to various existing theories
and models like Quantum Mechanics, QED, the Fermi model \etc;}
\item{the \emph{minimality}, \ie only observed and/or unavoidable objects
(fields and interactions) are involved;}
\item{the \emph{unitarity} which is a general condition for cross sections and
various transformations of fields related to the fact that any probability limited
from above by unity;}
\item{the \emph{renormalizability} is necessary for derivation of finite predictions
for observable quantities at the quantum level;}
\item{the \emph{gauge} principle for introduction of interactions (were possible).}
\end{itemize}

The main guiding principle is the \emph{symmetry} one. The SM possesses 
several different symmetries: \\
--- the Lorentz (and Poincar\'e) symmetry, \\
--- the CPT symmetry, \\
--- three gauge symmetries $SU(3)_C \otimes SU(2)_L \otimes U(1)_Y$, \\
--- the global $SU(2)_L\times SU(2)_R$ symmetry in the Higgs sector 
(it is broken spontaneously); \\
--- some other symmetries, like the one between three generation of fermions,
    the one that provides cancellation of axial anomalies \etc\\
In this context, one can mention also the conformal symmetry which is obviously
broken in the SM, but the mechanism of its breaking and the consequences are
very important for the model.

\section{Brief notes on Quantum field theory
\label{sec:QFT}}

The Standard Model is a model constructed within the local relativistic 
Quantum field theory. 
It means that the SM obeys the general QFT rules. We should keep in mind
that there are many other possible QFT models, and the SM is distinguished
between them mostly because of its successful experimental verifications
but also because of a number of its features like renormalizablity, unitarity, 
and cancellation of axial anomalies.
I assume that all students of the ESHEP school had courses on Quantum field theory. 
Here we will just remind several features of QFT which are important for further 
construction of the SM Lagrangian. 

As it was already mentioned, we are going to preserve the correspondence to
Quantum Mechanics (QM). Historically, QFT was developed on the base of QM,
in particular using the quantum oscillator ansatz. 
But by itself QFT can be considered as a more profound fundamental construction,
so one should be able to define this theory without referring to QM. In fact,
QFT can be formulated starting from the basic classification of fields as 
unitary irreducible representations of the Lorentz group.

Let us first of all fix the notation. We will work in the natural system
of units where the speed of light $c=1$ and the reduced Planck constant $\hbar = 1$.
The Lorentz indexes will be denoted by Greek letters, like $\mu = 0,1,2,3$;
$p_\mu$ is a four-momentum of a particle, 
$\mathbf{p} = (p_1,p_2,p_3)$ is a three-momentum, 
$p_0=E$ is the particle energy. 
The metric tensor of the Minkowsky space is chosen in the form
\begin{eqnarray} 
&& g_{\mu\nu} = \left( \begin{array}{cccc} 1 & 0 & 0 & 0 \\
0 & -1 & 0 & 0 \\ 0 & 0 & -1 & 0 \\ 0 & 0 & 0 & -1 
\end{array} \right), \quad g_{\mu\nu}p_\nu = p_\mu, \quad g_{\mu\mu} = 4. 
\end{eqnarray}
We will always assume summation over a Lorentz index if it is repeated twice:
$A_\mu B_\mu \equiv A_0 B_0 - A_1B_1 - A_2B_2 - A_3B_3$, where the metric tensor is used.
In particular, the scalar product of two four-vectors is defined as 
$pq=p_\mu q_\mu = p_0q_0 - p_1q_1 - p_2q_2 - p_3q_3$. 
It is a relativistic \emph{invariant}.

We will assume that there exist so-called asymptotic free final states
for particle-like excitation of quantum fields. These asymptotic states
will be associated with initial or final state (elementary) particles
which fly in a free space without interactions.
For such states we apply the \emph{on-mass-shell condition}
$p^2 = pp = p_0^2-\mathbf{p}^2 = E^2-\mathbf{p}^2= m^2$ where $m$ is the mass of the
particle. 

Now we will postulate the properties of fields that are required for
the construction of the SM.
A neutral scalar field can be defined as
\begin{eqnarray}
\varphi(x) = \frac{1}{(2\pi)^{3/2}}\int \frac{\mathrm{d}\mathbf{p}}{\sqrt{2p_0}}
\left(e^{-ipx}a^-(\mathbf{p})+e^{+ipx}a^+(\mathbf{p})\right),
\end{eqnarray}
where $a^\pm(\mathbf{p})$ are \emph{creation} and \emph{annihilation} operators.
Their commutation relations read
\begin{eqnarray}
&& [a^-(\mathbf{p}),a^+(\mathbf{p'})] \equiv a^-(\mathbf{p})a^+(\mathbf{p'})
-a^+(\mathbf{p}')a^-(\mathbf{p}) = \delta(\mathbf{p}-\mathbf{p'}), 
\nonumber \\ 
&& [a^-(\mathbf{p}),a^-(\mathbf{p'})] = [a^+(\mathbf{p}),a^+(\mathbf{p'})] = 0.
\end{eqnarray}
The field is a function of four-coordinate $x$ in the Minkowsky space. It behaves
as a plane wave in the whole space.
The Lagrangian\footnote{Actually it is a Lagrangian density.} 
for the neutral scalar field can be chosen in the form
\begin{eqnarray}
\mathcal{L}(x) = \frac{1}{2}(\partial_\mu\varphi\partial_\mu\varphi - m^2\varphi^2).
\end{eqnarray}
Note that it depends only on the field and its first derivative.
Variation of the action $A\equiv\int \mathrm{d}^4 \mathcal{L}(x)$ with respect to
variations of the field $\varphi\to\varphi+\delta\varphi$ according to the
\emph{ least action principle} gives
\begin{eqnarray}
\delta \int \mathrm{d} x \mathcal{L}(x) = \int \mathrm{d} x \biggl(
\frac{\partial \mathcal{L} }{\partial\varphi}\delta\varphi 
+ \frac{\partial \mathcal{L} }{\partial(\partial_\mu\varphi)}\delta(\partial_\mu\varphi)
\biggr) = 0.
\end{eqnarray}
Here we apply quite natural for QFT problems zero boundary conditions for the field and 
its derivative at infinity and get the well-known Klein--Fock--Gordon equation of motion
\begin{eqnarray}
(\partial_\mu^2+m^2)\varphi(x) = 0.
\end{eqnarray}
\emph{ EXERCISE: Check that the postulated above field $\varphi(x)$ satisfies the equation.}

Creation and annihilation operators act in the Fock space which consists of
\emph{ vacuum} ground state denoted as $|0\rangle$ and \emph{excitations} over it.
For the vacuum state we postulate
\begin{eqnarray}
a^-(\mathbf{p})|0\rangle = 0, \quad
\langle 0| a^+(\mathbf{p}) = 0, \quad
\langle 0|0\rangle = 1. 
\end{eqnarray}
Actually, $a^-(\mathbf{p})|0\rangle = 0\cdot|0\rangle $ but
the vacuum state can be dropped since finally all observable
quantities are proportional to $\langle 0|0\rangle$.
The field excitations are states of the form
\begin{eqnarray} 
|f\rangle =  \int \mathrm{d}\mathbf{p}\, f(\mathbf{p})a^+(\mathbf{p})|0\rangle, \quad
|g\rangle =  \int \mathrm{d}\mathbf{p} \mathrm{d}\mathbf{q}\, g(\mathbf{p},\mathbf{q})
a^+(\mathbf{p})a^+(\mathbf{q})|0\rangle,\quad \ldots 
\end{eqnarray}
The most simple excitation $a^+(\mathbf{p})|0\rangle\equiv|p\rangle$ is used to describe 
a single on-mass-shell particle with momentum $\mathbf{p}$. 
Then $a^+(\mathbf{p})a^+(\mathbf{q})|0\rangle$ is a two-particle state and so on. 
Because of the presence of modulating functions 
like $f(\mathbf{p})$ and $g(\mathbf{p},\mathbf{q})$, the Fock space is
infinite-dimensional. 
%Norm of a state in the Fock space is defined by
%\begin{eqnarray}
%||\, |f\rangle || \equiv  \langle f|f \rangle.
%\end{eqnarray}

\emph{ EXERCISES: 1) Find the norm $\langle p |p \rangle$; 2) check that
operator $\hat{N}=\int \mathrm{d}\mathbf{p}\, a^+(\mathbf{p})a^-(\mathbf{p})$ acts 
as a particle number operator.}

A charged scalar field is defined as
\begin{eqnarray}
\nonumber
&& \varphi(x) = \frac{1}{(2\pi)^{3/2}}\int \frac{\mathrm{d}\mathbf{p}}{\sqrt{2p_0}}
\left(e^{-ipx}a^-(\mathbf{p})+e^{+ipx}b^+(\mathbf{p})\right),
\\ \nonumber
&& \varphi^*(x) = \frac{1}{(2\pi)^{3/2}}\int \frac{\mathrm{d}\mathbf{p}}{\sqrt{2p_0}}
\left(e^{-ipx}b^-(\mathbf{p})+e^{+ipx}a^+(\mathbf{p})\right),
\\ \nonumber
&& [a^-(\mathbf{p}),a^+(\mathbf{p'})] = [b^-(\mathbf{p}),b^+(\mathbf{p'})] = \delta(\mathbf{p}-\mathbf{p'}), 
\quad
[a^\pm,b^\pm] = 0,
\end{eqnarray}
where operators $a^\pm(\mathbf{p})$ create and annihilate particles,
while operators $b^\pm(\mathbf{p})$ are used for the same purpose for \emph{anti-particles}.
Note that the choice of what is particle and what is anti-particle is arbitrary here.
The corresponding Lagrangian reads
\begin{eqnarray} \label{L_phi_c}
\mathcal{L}(\phi,\phi^*) = \partial_\mu\varphi^* \partial_\mu\varphi - m^2\varphi^*\varphi.
\end{eqnarray}
Note that $\varphi$ and $\varphi^*$ are related by a generalized complex conjugation which
involves operator transformations: $(a^\pm)^* = a^\mp$ and  $(b^\pm)^* = b^\mp$.
It is worth to note also that $\varphi$ and $\varphi^*$ are not ``a particle and 
an anti-particle''.

A massive charged vector field (remind $W^\pm$ bosons) is defined as
\begin{eqnarray} \nonumber
&& U_\mu(x) = \frac{1}{(2\pi)^{3/2}}\int \frac{\mathrm{d}\mathbf{p}}{\sqrt{2p_0}}
\sum_{n=1,2,3}e^n_\mu(\mathbf{p})\left(e^{-ipx}a^-_n(\mathbf{p})+e^{+ipx}b^+_n(\mathbf{p})\right),
\\ \nonumber
&& U_\mu^*(x) = \frac{1}{(2\pi)^{3/2}}\int \frac{\mathrm{d}\mathbf{p}}{\sqrt{2p_0}}
\sum_{n=1,2,3}e^n_\mu(\mathbf{p})\left(e^{-ipx}b^-_n(\mathbf{p})+e^{+ipx}a^+_n(\mathbf{p})\right),
\\ \nonumber
&& [a^-_n(\mathbf{p}),a^+_l(\mathbf{p'})] = [b_n^-(\mathbf{p}),b_l^+(\mathbf{p'})] =
\delta_{nl} \delta(\mathbf{p}-\mathbf{p'}), 
\quad
[a^\pm,b^\pm] = 0.
\end{eqnarray}

For \emph{polarization vectors} $e^n_\mu(\mathbf{p})$ the following conditions 
are applied:
\begin{eqnarray}
e^n_\mu(\mathbf{p})e^l_\mu(\mathbf{p}) = -\delta_{nl}, 
\qquad p_\mu e^n_\mu(\mathbf{p}) = 0. 
\end{eqnarray}
\emph{ EXERCISE: Using the above orthogonality conditions, show that}
\begin{eqnarray}
\sum_{n=1,2,3}e^n_\mu(\mathbf{p})e^n_\nu(\mathbf{p})
= - \biggl( g_{\mu\nu} - \frac{p_\mu p_\nu}{m^2} \biggr).
\end{eqnarray}

The Lagrangian for a massive charged vector field takes the form
\begin{eqnarray} 
\mathcal{L} = - \frac{1}{2}\biggl(\partial_\mu U^*_\nu - \partial_\nu U^*_\mu\biggr) 
\biggl(\partial_\mu U_\nu - \partial_\nu U_\mu\biggr) + m^2U^*_\mu U_\mu. 
\end{eqnarray}
The corresponding Euler--Lagrange equation reads
\begin{eqnarray} \nonumber
\partial_\nu(\partial_\mu U_\nu - \partial_\nu U_\mu) + m^2U_\mu = 0.
\end{eqnarray}
\emph{ EXERCISE: Using the above equation, show that $\partial_\nu U_\nu(x)=0$,
\ie derive the \emph{Lorentz condition}}.
Note that the Lorentz condition removes from the field one of four independent 
degrees of freedom (components).

A massless neutral vector field (a photon) is defined as
\begin{eqnarray} 
&& A_\mu(x) = \frac{1}{(2\pi)^{3/2}}\int \frac{\mathrm{d}\mathbf{p}}{\sqrt{2p_0}}
e^\lambda_\mu(\mathbf{p})\left(e^{-ipx}a^-_\lambda(\mathbf{p})
+e^{+ipx}a^+_\lambda(\mathbf{p})\right),
\\ \nonumber
&& [a^-_\lambda(\mathbf{p}),a^+_\nu(\mathbf{p'})]
= - g_{\lambda\nu} \delta(\mathbf{p}-\mathbf{p'}) 
\qquad
e^\lambda_\mu(\mathbf{p})e^\lambda_\nu(\mathbf{p}) = g_{\mu\nu},
\qquad
e^\lambda_\mu(\mathbf{p})e^\nu_\mu(\mathbf{p}) = g_{\lambda\nu}.
\end{eqnarray}
Formally this field has four polarizations, but only two of them correspond
to physical degrees of freedom. 
The corresponding Lagrangian reads
\begin{eqnarray} \label{L_A}
\mathcal{L} = - \frac{1}{4}F_{\mu\nu}F_{\mu\nu},\qquad 
F_{\mu\nu}\equiv  \partial_\mu A_\nu - \partial_\nu A_\mu.
\end{eqnarray}

A Dirac fermion field is defined as
\begin{eqnarray} \label{spinor_field}
&& \Psi(x) = \frac{1}{(2\pi)^{3/2}}\int \frac{\mathrm{d}\mathbf{p}}{\sqrt{2p_0}}
\sum_{r=1,2}\left(e^{-ipx}a^-_r(\mathbf{p})u_r(\mathbf{p})
+e^{+ipx}b^+_r(\mathbf{p})v_r(\mathbf{p})\right),
\\ \nonumber
&& \overline{\Psi}(x) = \frac{1}{(2\pi)^{3/2}}\int \frac{\mathrm{d}\mathbf{p}}{\sqrt{2p_0}}
\sum_{r=1,2}\left(e^{-ipx}b^-_r(\mathbf{p})\bar{v}_r(\mathbf{p})
+e^{+ipx}a^+_r(\mathbf{p})\bar{u}_r(\mathbf{p})\right),
\\ \nonumber
&& [a^-_r(\mathbf{p}),a^+_s(\mathbf{p'})]_{+}
= [b^-_r(\mathbf{p}),b^+_s(\mathbf{p'})]_{+}
= \delta_{rs} \delta(\mathbf{p}-\mathbf{p'}), 
\\ \nonumber
&& [a^+_r(\mathbf{p}),a^+_s(\mathbf{p'})]_+
= [a^-_r(\mathbf{p}),b^+_s(\mathbf{p'})]_+
= \ldots = 0.
\end{eqnarray}
\emph{ EXERCISE: Show that $a^+_r(\mathbf{p})a^+_r(\mathbf{p})=0$,
\ie verify the \emph{Pauli principle}}.

Here, $u_r$, $u_r$, $\bar{u}_r$, and $\bar{v}_r$ are four-component spinors,
so $\Psi(x)\equiv\{\Psi_\alpha(x)\}$ is a four-vector column, $\alpha = 1,2,3,4$,
and $\overline{\Psi}(x)$ is a four-vector row, 
\begin{eqnarray} \nonumber
\bar{u}u = \sum_{\alpha=1}^4 \bar{u}_\alpha u_\alpha = 
\sum_{\alpha=1}^4  u_\alpha \bar{u}_\alpha =  \mathrm{Tr}(u\bar{u}).
\end{eqnarray}
Spinors are solutions of the (Dirac) equations:
\begin{eqnarray} \label{Dir_eq}
&& (\hat{p}-m)u_r(\mathbf{p})=0, \qquad \bar{u}_r(\mathbf{p})(\hat{p}-m)=0,
\\ \nonumber
&& (\hat{p}+m)v_r(\mathbf{p})=0, \qquad \bar{v}_r(\mathbf{p})(\hat{p}+m)=0,
\\ \nonumber
&& \hat{p} \equiv p_\mu \gamma_\mu = p_0\gamma_0-p_1\gamma_1-p_2\gamma_2-p_3\gamma_3,
\qquad m \equiv m \mathbf{1},
\end{eqnarray}
where $\mathbf{1}$ is the unit four-by-four matrix.
For the solutions of the above equations we impose the normalization conditions 
\begin{eqnarray} \nonumber
\bar{u}_r(\mathbf{p})u_s(\mathbf{p})
= - \bar{v}_r(\mathbf{p})v_s(\mathbf{p})=2m\delta_{rs}.
\end{eqnarray}

The gamma matrixes (should) satisfy the commutation condition
\begin{eqnarray} \nonumber
[\gamma_\mu,\gamma_\nu]_+ = 2g_{\mu\nu} \mathbf{1} \quad
\Rightarrow \quad \gamma_0^2 = \mathbf{1}, \quad
\gamma_1^2=\gamma_2^2=\gamma_3^2=-\mathbf{1}
\end{eqnarray}
and the condition of Hermitian conjugation 
\begin{eqnarray} \nonumber
\gamma_\mu^\dagger = \gamma_0\gamma_\mu\gamma_0.
\end{eqnarray}
The latter leads to the rule of the \emph{Dirac conjugation}:
\begin{eqnarray} \label{Dirac_conj}
\overline{\Psi} = \Psi^\dagger \gamma_0, \quad
\bar{u} = u^\dagger\gamma_0, \quad \bar{v} = v^\dagger\gamma_0. 
\end{eqnarray}
\emph{ EXERCISE: Show that the Dirac conjugation rule is consistent with
the set of Dirac equations~(\ref{Dir_eq})}.

Note that explicit expressions for gamma matrixes are not unique,
but they are not necessary for construction of observables, 
\emph{ QUESTION: Why is that so?}
The most common representations of gamma matrixes are so-called 
Dirac's (standard) and Weyl's (spinor) ones. 

Two values of index $r$ in Eq.~(\ref{spinor_field}) correspond to two
independent degrees of freedom for each spinor in other words to
two independent solutions of the Dirac equations. In most cases these 
two degrees of freedom
can be treated as two polarization states like 'spin up' and 'spin down'.  
But in the Standard Model, there is one special choice of the basis for
spinors, namely we will distinguish \emph{Left} (L) and \emph{Right} (R) 
chiral states of spinors.
By definition,
\begin{eqnarray} \label{left_right}
\Psi_L \equiv P_L\Psi, \quad \Psi_R \equiv P_R\Psi, \qquad 
P_{L,R} \equiv \frac{\mathbf{1}\ -,+\ \gamma_5}{2}, \qquad \Psi=\Psi_L + \Psi_R.
\end{eqnarray} 
Here $\gamma_5 \equiv i\gamma_0\gamma_1\gamma_2\gamma_3$, this gamma-matrix
has the properties
\begin{eqnarray}
[\gamma_\mu,\gamma_5]_+ = 0, \qquad
\gamma_5^2=\mathbf{1}, \quad \gamma_5^\dagger = \gamma_5.
\end{eqnarray}
As can be seen from Eq.~(\ref{left_right}), $P_{L,R}$ form a complete set of orthogonal
projection operators,
\begin{eqnarray}
P_{L,R}^2 = P_{L,R} \qquad P_LP_R=P_RP_L = 0, \qquad P_L+P_R=1.
\end{eqnarray}
The sign before $\gamma_5$ in the definition of the projection operators
in Eq.~(\ref{left_right}) corresponds to the standard representation
of gamma matrixes\footnote{In the spinor representation the sign is opposite:
$P_L \equiv (1+\gamma_5)/2$ and $P_R\equiv (1-\gamma_5)/2$.}.
The Dirac conjugation~(\ref{Dirac_conj}) of left and right spinors gives
\begin{eqnarray} \nonumber
\overline{\Psi}_L \equiv \overline{\Psi} \frac{1+\gamma_5}{2}, \qquad
\overline{\Psi}_R \equiv \overline{\Psi} \frac{1-\gamma_5}{2}.
\end{eqnarray}
Note that the definition of the left and right chiral states was done
without referring to spin projections (helicity states). In fact, these
are different ways to select a basis. Helicity and chirality states
can be identified to each other only for massless fermions.

Remind some properties of gamma matrixes
\begin{eqnarray} \nonumber
&& \mathrm{Tr}\gamma_\mu = \mathrm{Tr}\gamma_5 =0, \quad
\mathrm{Tr}\gamma_\mu\gamma_\nu = 4g_{\mu\nu}, \quad
\mathrm{Tr}\gamma_5\gamma_\mu\gamma_\nu = 0,
\\ \nonumber
&& \mathrm{Tr}\gamma_\mu\gamma_\nu\gamma_\alpha\gamma_\beta = 4(g_{\mu\nu}g_{\alpha\beta}
-g_{\mu\alpha}g_{\nu\beta}+g_{\mu\beta}g_{\nu\alpha}), \qquad 
\mathrm{Tr}\gamma_5\gamma_\mu\gamma_\nu\gamma_\alpha\gamma_\beta =
- 4i\varepsilon_{\mu\nu\alpha\beta}.
\end{eqnarray}
The equations for $u$ and $v$ are chosen so that we get the conventional \emph{Dirac equations}
\begin{eqnarray} \nonumber
(i\gamma_\mu\partial_\mu-m)\Psi(x) = 0, \qquad
i\partial_\mu\overline{\Psi}(x)\gamma_\mu+m\overline{\Psi}(x) = 0.
\end{eqnarray}

These equations follow also from the Lagrangian
\begin{eqnarray} \nonumber
\mathcal{L} 
= \frac{i}{2}\biggl[\overline{\Psi}\gamma_\mu(\partial_\mu\Psi)
- (\partial_\mu\overline{\Psi})\gamma_\mu\Psi\biggr] 
- m\overline{\Psi}\Psi 
\equiv i\overline{\Psi}\gamma_\mu\partial_\mu\Psi
- m\overline{\Psi}\Psi. 
\end{eqnarray}
Note that the right-hand side is a short notation for the explicit Lagrangian
which is given in the middle.

In QFT, Lagrangians (Hamiltonians) should be Hermitian: 
$\mathcal{L}^\dagger = \mathcal{L}$. 
\emph{ QUESTION: What kind of problems one can have with a non-Hermitian Hamiltonian?} 

Up to now we considered only \emph{free} non-interacting fields.
Studies of transitions between free states is the main task of QFT\footnote{
Collective, nonperturbative effects, bound states \etc are also of
interest, but that goes beyond the scope of these lectures.}.

Let us postulate the transition \emph{amplitude} (matrix element) 
$\mathcal{M}$ of a physical process:
\begin{eqnarray} \label{M_element}
\mathcal{M} \equiv \langle out| S | in \rangle, \qquad
S \equiv T \exp\biggl(i\int \mathrm{d} x \mathcal{L}_I(\varphi(x))\biggr). 
\end{eqnarray}
Here $S$ is the so-called S-matrix which is the general evolution operator 
of quantum states.
Letter $T$ means the \emph{time ordering} operator, it will be discussed a bit later.
The initial and final states are
\begin{eqnarray}
|in\rangle = a^+(\mathbf{p}_1) \ldots a^+(\mathbf{p}_s) |0\rangle, \qquad 
|out\rangle = a^+(\mathbf{p'}_1) \ldots a^+(\mathbf{p'}_r) |0\rangle.
\end{eqnarray}

The differential probability to evolve from $|in\rangle$ to $|out\rangle$ is
\begin{eqnarray} \nonumber
\mathrm{d} w = (2\pi)^4\delta(\sum {p'}_i)\frac{n_1\ldots n_s}{2E_1\ldots E_s}
|\mathcal{M}|^2 \prod_{j=1}^{r} \frac{\mathrm{d}\mathbf{p'}_j}{(2\pi)^32{E'}_j}.
\end{eqnarray}
Here $n_i$ is the particle number density of $i^\mathrm{th}$ particle beam.

Nontrivial transitions happen due to interactions of fields. QFT prefers
dealing with \emph{local} interactions 
$\Rightarrow\ \ \mathcal{L}_I = \mathcal{L}_I(\varphi(x))$.
By 'local' we mean that all interaction terms in the Lagrangian are constructed
as products of fields (or their first derivatives) taken at the same space-time
coordinate.

Here are some examples of interaction Lagrangians:
\begin{eqnarray} \nonumber
&& g\varphi^3(x), \qquad h\varphi^4(x), \qquad y\varphi(x)\overline{\Psi}(x)\Psi(x), 
\\ \nonumber
&& e \overline{\Psi}(x)\gamma_\mu\Psi(x)A_\mu(x), \qquad
G \overline{\Psi}_1(x)\gamma_\mu\Psi_1(x)\cdot \overline{\Psi}_2(x)\gamma_\mu\Psi_2(x).
\end{eqnarray} 
\emph{ IMPORTANT: Always keep in mind the \emph{dimension} of your objects!}
The reference unit is the dimension of energy (mass):
\begin{eqnarray}
[E] =[m] = 1\quad \Rightarrow\quad [p] = 1,\qquad [x] = -1.
\end{eqnarray}
An action should be dimensionless 
\begin{eqnarray}
\biggl[\int \mathrm{d} x \mathcal{L}(x)\biggr] = 0\ \Rightarrow \ [\mathcal{L}] = 4. 
\end{eqnarray}

\emph{ EXERCISE: Show that $[\varphi]=[A_\mu]=1$ and $[\Psi] = 3/2$. Find the
dimensions of the coupling constants $g$, $h$, $y$, $e$, and $G$ in the examples above.}

By definition the time ordering operator acts as follows:
\begin{eqnarray}
T\ A_1(x_1) \ldots A_n(x_n) = (-1)^l A_{i_1}(x_{i_1})\ldots A_{i_n}(x_{i_n})\ \ 
\mathrm{with} \ \ x^0_{i_1}>\ldots >x^0_{i_n}, 
\end{eqnarray}
where $l$ is the number of fermion field permutations.

The perturbative expansion of the $S$ matrix exponent~(\ref{M_element}) 
leads to terms like
\begin{eqnarray} \nonumber
\frac{i^ng^n}{n!}\langle 0| a^-(\mathbf{p'}_1)\ldots a^-(\mathbf{p'}_r) 
\int \mathrm{d} x_1\ldots \mathrm{d} x_n T \varphi^3(x_1)\ldots \varphi^3(x_n)
a^+(\mathbf{p}_1)\ldots a^+(\mathbf{p}_s) |0\rangle.
\end{eqnarray}

Remind that fields $\varphi$ also contain creation and annihilation operators.
By permutation of operators 
$a^-(\mathbf{p})a^+(\mathbf{p'}) = a^+(\mathbf{p'})a^-(\mathbf{p}) 
+ \delta(\mathbf{p}-\mathbf{p'})$
we move $a^-$ to the right and $a^+$ to the left.
At the end we get either zero because $a^-|0\rangle =0$ and $\rangle 0|a^+ =0$,
or some finite terms proportional to $\langle 0|0\rangle =1$.

\emph{ EXERCISE: Show that $[a^-(\mathbf{p}),\varphi(x)] = \frac{e^{ipx}}{(2\pi)^{3/2}\sqrt{2p_0}}$
and $[a_r^-(\mathbf{p}),\overline{\Psi}(x)]_+ 
= \frac{e^{ipx}\bar{u}_r(\mathbf{p})}{(2\pi)^{3/2}\sqrt{2p_0}}$.}

By definition the \emph{causal Green function} is given by 
\begin{eqnarray} \label{D_c_def}
\langle 0|T\varphi(x)\varphi(y)|0\rangle \equiv - i D^c(x-y). 
\end{eqnarray}
It is a building block for construction of amplitudes.
One can show (see textbooks) that
\begin{eqnarray}
(\partial^2+m^2)D^c(x) = \delta(x), 
\end{eqnarray}
so that $D^c$ is the Green function of the Klein--Fock--Gordon operator,
\begin{eqnarray}
D^c(x) = \frac{-1}{(2\pi)^4}\int\frac{\mathrm{d} p\; e^{-ipx}}{p^2-m^2+i0},
\end{eqnarray}
where $+i0$ is an infinitesimally small imaginary quantity
which shifts the poles of the Green function from the real axis in the complex plane.
The sign of this quantity is chosen to fulfil the requirement 
of the time ordering operation in Eq.~(\ref{D_c_def}).

For other fields we have
\begin{eqnarray} \nonumber
&& \langle 0 | T\; \Psi(x)\overline{\Psi}(y) | 0 \rangle 
= \frac{i}{(2\pi)^4}\int\frac{\mathrm{d} p\; e^{-ip(x-y)}(\hat{p}+m)}{p^2-m^2 +i0},
\\ \label{eq:vector_prop}
&& \langle 0 | T\; U_\mu(x)U^*_\nu(y) | 0 \rangle 
= \frac{-i}{(2\pi)^4}\int\frac{\mathrm{d} p\; e^{-ip(x-y)}(g_{\mu\nu}-p_\mu p_\nu/m^2)}{p^2-m^2+i0},
\\ \nonumber
&& \langle 0 | T\; A_\mu(x)A_\nu(y) | 0 \rangle 
= \frac{-i}{(2\pi)^4}\int\frac{\mathrm{d} p\; e^{-ip(x-y)}g_{\mu\nu}}{p^2+i0}.
\end{eqnarray}
The Wick theorem states that for any combinations of fields
\begin{eqnarray}
T\; A_1\ldots A_n \equiv \sum (-1)^l \langle0|TA_{i_1}A_{i_2}|0\rangle
\ldots \langle0|TA_{i_{k-1}}A_{i_k}|0\rangle\; :A_{i_{k+1}}\ldots A_{i_n}: 
\end{eqnarray}
The sum is taken over all possible ways to pair the fields.

The \emph{normal ordering} operation acts as
\begin{eqnarray} 
:a_1^- a_2^+ a_3^- a_4^- a_5^+ a_6^- a_7^+: \ =\ (-1)^l a_2^+ a_5^+ a_7^+ a_1^- a_3^- a_4^- a_6^- 
\end{eqnarray}
so that all annihilation operators go to the right and all creation operators go to the left.
The number of fermion operator permutations $l$ provides the factor $(-1)^l$. 

Using the Wick theorem we construct the \emph{Feynman rules} for simple $g\phi^3$ and $h\varphi^4$
interactions. But for the case of gauge interactions we need something more as we will see below.

It appears that symmetries play a crucial role in the QFT. 
There are two major types of symmetries in the SM: \emph{global} and \emph{local} ones.
By a global symmetry we mean invariance of a Lagrangian and observables with respect to certain
transformations of coordinates and/or fields if the transformations are the same 
in each space-time point. 
If the transformations do depend on coordinates, the corresponding symmetry is called local.

The \textbf{1st Noether (N\"other) theorem}: \\
\emph{If an action is invariant with respect to transformations of a global Lie group $G_r$ 
with $r$ parameters, then there are $r$ linearly independent combinations of Lagrange 
derivatives which become complete divergences; and vice versa. }

If the field satisfies the Euler--Lagrange equations, then $\mathrm{div} J=\nabla J = 0$,
\ie the \emph{Noether currents} are conserved.
Integration of those divergences over a 3-dimensional volume (with certain boundary conditions)
leads to $r$ \emph{conserved charges}.  
Remind that conservation of the electric charge in QED is related to the global $U(1)$ symmetry
of this model, and that Poincar\'e symmetries lead to conservation of energy, momentum, 
and angular momentum.  

Much more involved and actually important for us is the
\textbf{2nd Noether theorem}: \\
\emph{If the action is invariant with respect to the infinite-dimensional $r$-parametric 
group $G_{\infty,r}$ with derivatives up to the $k^{\mathrm{th}}$ order, then there are 
$r$ independent relations between Lagrange derivatives and derivatives of them up to
the $k^{\mathrm{th}}$ order; and vice versa. }

The importance of the second theorem is justified by the fact that gauge groups
(and also the general coordinate transformation in Einstein's gravitational theory)
are infinite-dimensional groups.
The 2nd Noether theorem provides $r$ conditions on the fields which are additional 
to the standard Euler--Lagrange equations. These conditions should be used to exclude 
\emph{ double counting} of physically \emph{equivalent} field configurations.

\subsection{Gauge symmetries}

Let us start the discussion of local gauge symmetries with 
Quantum Electrodynamics (QED). 
The free Lagrangians for electrons and photons
\begin{eqnarray}
\mathcal{L}_0(\Psi) = i\overline{\Psi} \gamma_\mu\partial_\mu \Psi 
- m\overline{\Psi}\Psi, \quad 
\mathcal{L}_0(A) = -\frac{1}{4}F_{\mu\nu}F_{\mu\nu} 
\end{eqnarray}
are invariant with respect to the \emph{global} $U(1)$ transformations  
\begin{eqnarray} \label{global_U1}
\Psi(x) \to \exp(ie\theta)\Psi(x), \quad 
\overline{\Psi}(x) \to \exp(-ie\theta)\overline\Psi(x), \quad
A_\mu(x) \to A_\mu(x). 
\end{eqnarray}

One can note that $F_{\mu\nu}$ is invariant also with respect to 
\emph{local} transformations $A_\mu(x) \to A_\mu(x) + \partial_\mu \omega(x)$,
where $\omega(x)$ is an arbitrary (differentiable) function.
For fermions the corresponding transformations are
\begin{eqnarray}
\Psi(x) \to \exp(ie\omega(x))\Psi(x), \quad 
\overline{\Psi}(x) \to \exp(-ie\omega(x))\overline\Psi(x), 
\end{eqnarray}
\ie where the global constant angle $\theta$ in Eq.~(\ref{global_U1})
is substituted by a local function $\omega(x)$ which varies from one space-time
point to another.

The question is how to make the fermion Lagrangian being also invariant
with respect to the local transformations?
The answer is to introduce the so-called \emph{covariant derivative:}
\begin{eqnarray}
\partial_\mu \to D_\mu, \qquad D_\mu\Psi \equiv (\partial_\mu - ieA_\mu)\Psi,\quad
D_\mu\overline{\Psi} \equiv (\partial_\mu + ieA_\mu)\overline{\Psi}. 
\end{eqnarray}
Then we get the QED Lagrangian
\begin{eqnarray} \nonumber 
\mathcal{L}_{\mathrm{QED}} &=& - \frac{1}{4}F_{\mu\nu} F_{\mu\nu} 
+ i\overline{\Psi}\gamma_\mu D_\mu \Psi - m\overline{\Psi}\Psi
\\ \nonumber 
&=& - \frac{1}{4}F_{\mu\nu} F_{\mu\nu} + i\overline{\Psi}\gamma_\mu \partial_\mu \Psi 
- m\overline{\Psi}\Psi + e\overline{\Psi} \gamma_\mu \Psi A_\mu, 
\end{eqnarray}
where the last term describes interaction of electrons and positrons with photons.
The most important point here is that the structure of the interaction term is 
completely fixed by the gauge symmetry. Nevertheless, there is one specific
feature of the abelian $U(1)$ case, namely the values of electric charges
(coupling constants) can be different for different fermions \eg for up and down quarks.  

%One can note that local transformations are more natural because of causality
%of Special relativity. 

\emph{ EXERCISES: 1) Check the covariance:
$ D_\mu\Psi \to e^{ie\omega(x)} (D_\mu\Psi)$; 
2) construct the Lagrangian of scalar QED (use Eqs.~(\ref{L_phi_c}) and (\ref{L_A})).}

Let's look now again at the free photon Lagrangian
\begin{eqnarray} \nonumber
&& \mathcal{L}_0(A) = - \frac{1}{4}(\partial_\mu A_\nu - \partial_\nu A_\mu )^2 
= - \frac{1}{2}A_\nu K_{\mu\nu} A_\nu, 
\\ \nonumber
&& K_{\mu\nu} = g_{\mu\nu}\partial^2 - \partial_\mu\partial_\nu \quad
\Rightarrow K_{\mu\nu}(p) = p_\mu p_\nu - g_{\mu\nu} p^2. 
\end{eqnarray}
Operator $K_{\mu\nu}(p)$ has zero modes (since $p_\mu K_{\mu\nu}=0)$, so it is not
invertable. Definition of the photon propagator within the \emph{functional
integral formalism} becomes impossible. The reason is the unresolved symmetry. 
The solution is to introduce a \emph{gauge fixing term} into the Lagrangian:
\begin{eqnarray} \nonumber
&& \mathcal{L}_0(A) = - \frac{1}{4}F_{\mu\nu}F_{\mu\nu} - \frac{1}{2\alpha}(\partial_\mu A_\mu)^2 
\quad \Rightarrow
\\ \nonumber
&& \langle 0 | T\; A_\mu(x)A_\nu(y) | 0 \rangle 
= \frac{-i}{(2\pi)^4}\int\! \mathrm{d} p\; e^{-ip(x-y)} \frac{g_{\mu\nu}+(\alpha-1)p_\mu p_\nu/p^2}{p^2 +i0}.
\end{eqnarray}
It is very important that physical quantities do not depend on the value of $\alpha$.

Let us briefly discuss the features of non-abelian Gauge symmetries which
will be also used in the construction of the SM.
Transformations for a \emph{non-abelian} case read
\begin{eqnarray} \nonumber
&& \Psi_i \to \exp{ig\omega^a t_{ij}^a}\Psi_j, \qquad [t^a,t^b] = if^{abc}t^c,
\\ \nonumber
&& B_\mu^a \to B_\mu^a + \partial_\mu \omega^a + g f^{abc} B_\mu^b \omega^c, \quad
F^a_{\mu\nu}\equiv \partial_\mu B_\nu^a-\partial_\nu B_\mu^a + g f^{abc} B_\mu^b B_\nu^c,
\end{eqnarray}
where $t^a$ are the group generators, $f^{abc}$ are the structure constants
(see details in the lectures on QCD). 

We introduce the covariant derivative
\begin{eqnarray} \nonumber
\partial_\mu\Psi \to  D_\mu\Psi \equiv (\partial_\mu -igB_\mu^a t^a)\Psi 
\end{eqnarray}
and get
\begin{eqnarray} \nonumber
\mathcal{L}(\Psi,B) &=& i\overline{\Psi}\gamma_\mu D_\mu\Psi +\mathcal{L}(B),
\\ \nonumber
\mathcal{L}(B) &=& - \frac{1}{4}F_{\mu\nu}^aF_{\mu\nu}^a 
- \frac{1}{2\alpha}(\partial_\mu B_\mu^a)^2 
= - \frac{1}{4}\left(\partial_\mu B_\nu^a-\partial_\nu B_\mu^a \right)^2
- \frac{1}{2\alpha}(\partial_\mu B_\mu^a)^2 
\\ \nonumber
&-& \frac{g}{2}f^{abc}\left(\partial_\mu B_\nu^a-\partial_\nu B_\mu^a \right)B_\mu^b B_\nu^c
- \frac{g^2}{4}f^{abc}f^{ade}B_\mu^b B_\nu^c B_\mu^d B_\nu^e.
\end{eqnarray}
Note that $\mathcal{L}(B)$ contains self-interactions and can not be treated
as a 'free Lagrangian'.
There is no any mass term for the gauge field in the Lagrangian, $m_B\equiv 0$,
because such a term would be not gauge-invariant.
It is worth to note that the non-abelian charge $g$ is \emph{universal},
\ie it is the same for all fields which are transformed by the
given group.

Exclusion of double-counting due to the physical equivalence
of the field configurations related to each other by non-abelian 
gauge transformations is nontrivial. Functional integration 
over those identical configurations (or application of the BRST method)
leads to the appearance of the so-called \emph{Faddeev--Popov ghosts}: 
\begin{eqnarray} \nonumber
&&\!\!\! \mathcal{L}(\Psi,B) \to  \mathcal{L}(\Psi,B) + \mathcal{L}_{gh},
\\ 
&&\!\!\! \mathcal{L}_{gh} = - \partial_\mu \bar{c}^a \partial_\mu c^a
+ g f^{acb} \bar{c}^a B_\mu^c \partial_\mu c^a 
=  - \partial_\mu \bar{c}^a \partial_\mu c^a
- g f^{acb} \partial_\mu\bar{c}^a B_\mu^c  c^a,
\end{eqnarray}
where $c$ and $\bar{c}$ are ghost fields, they are fermion-like states with a 
boson-like kinetic term.
Keep in mind that Faddeev--Popov ghosts are \emph{fictitious} particles. 
In the Feynman rules they (should) appear only as virtual states in propagators
but not in the initial and final asymptotic states.  
Formally, ghosts can be found also in QED, but they are non-interacting since 
$f^{abc}=0$ there, and can be totally omitted.

\subsection{Regularization and renormalization}

Higher-order terms in the perturbative series contain loop integrals
which can be ultraviolet (UV) divergent, \eg
\begin{eqnarray} 
I_2 \equiv \int\frac{\mathrm{d}^4p}{(p^2+i0)((k-p)^2+i0)} \sim
\int\frac{|p|^3\; \mathrm{d}|p|}{|p|^4} \sim \ln{\infty}.
\end{eqnarray}
Introduction of an upper cut-off $M$ on the integration variable
leads to a finite, \ie \emph{regularized} value of the integral:
\begin{eqnarray} 
I_2^{\mathrm{cut-off}}\! = i\pi^2\biggl(\ln\frac{M^2}{k^2} + 1\! \biggr) 
+ \mathcal{O}\!\left(\frac{k^2}{M^2}\right)
= i\pi^2\biggl(\ln\frac{M^2}{\mu^2} 
- \ln\frac{k^2}{\mu^2} + 1\! \biggr) + \mathcal{O}\!\left(\frac{k^2}{M^2}\right).
\end{eqnarray}
Another possibility is the \emph{dimensional regularization} 
where $dim=4 \to dim = 4-2\varepsilon$
\begin{eqnarray} 
I_2^{\mathrm{dim.reg.}} = \mu^{2\varepsilon} \int\frac{\mathrm{d}^{4-2\varepsilon}p}{(p^2+i0)((k-p)^2+i0)} 
= i\pi^2\biggl(\frac{1}{\varepsilon} 
- \ln\frac{k^2}{\mu^2} + 2 \biggr) + \mathcal{O}\left(\varepsilon\right).
\end{eqnarray}
Here the divergence is parameterized by the $\varepsilon^{-1}$ term.
The origin of UV divergences is the \emph{locality} of interactions in QFT.

Let's consider a three-point (vertex) function in the $g\phi^3$ model, it looks like
\begin{eqnarray} \nonumber
&& G = \int \mathrm{d} x\, \mathrm{d} y\, \mathrm{d} z\; \varphi(x) \varphi(y) \varphi(z) F(x,y,z),
\\  \nonumber
&& F^{\mathrm{dim.reg.}} = \frac{A}{\varepsilon}\delta(y-x)\delta(z-x) + \ldots
\end{eqnarray}
\emph{ IMPORTANT:} It means that UV-divergent terms are \emph{local} 
(here because of the delta-functions). 

\emph{A QFT model is called \textbf{renormalizable} if all UV-divergent
terms are of the type of the ones existing in the original (semi)classical Lagrangian.
Otherwise the model is \textbf{non-renormalizable}. }

\emph{ EXAMPLES:} \\
a) renormalizable models: QED,\ QCD,\ the SM [proved by 't~Hooft \& Veltman],
$h\varphi^4$,\ $g\varphi^3$; \\
b) non-renormalizable models: the Fermi model with 
$\mathcal{L}\sim G(\overline{\Psi}\gamma_\mu\Psi)^2$ and General Relativity.

It can be shown that models with dimensionful $([G]<0)$ coupling constants are 
non-renormalizable.

In renormalizable models all UV divergences can be \emph{subtracted} 
from amplitudes and shifted into \emph{counter terms} in $\mathcal{L}$.
In this way each term in $\mathcal{L}$ gets a \emph{renormalization
constant}. For the model describing a scalar field with the $\varphi^4$ 
self-interaction we get
\begin{eqnarray} \nonumber
&& \mathcal{L} = \frac{Z_2}{2}(\partial \varphi)^2
- \frac{Z_m m^2}{2}\varphi^2 + Z_4h\varphi^4
= \frac{1}{2}(\partial \varphi_B)^2
- \frac{m^2_B}{2}\varphi^2 + h_B \varphi^4,
\end{eqnarray}
where $\varphi_B = \sqrt{Z_2}\varphi$,\  $m_B^2 = m^2Z_MZ_2^{-1}$,\
$h_B=hZ_4 Z_2^{-2}$ are so-called \emph{bare} field, mass, and charge,
\begin{eqnarray} \nonumber
Z_i(h,\varepsilon) = 1 + \frac{Ah}{\varepsilon} + \frac{Bh^2}{\varepsilon^2}
 + \frac{Ch^2}{\varepsilon} + \mathcal{O}(h^3). 
\end{eqnarray}
Remonrmalization constants are chosen in such a way that divergences in
amplitudes are \emph{cancelled out} with divergences in $Z_i$. 
By construction, that happens order by order. 

R.~Feynman said once: \emph{``I think that the renormalization theory is simply 
a way to sweep the difficulties of the divergences of electrodynamics under the rug.''} 
Physicists are still not fully satisfied by the renormalization procedure, 
but the method has been verified in many models. Moreover, renormalizable models
including the SM appear to be the most successful ones in the description of phenomenology. 
For these reasons we say now that renormalization is the general feature
of physical theories.

Physical results should not depend on the auxiliary scale $\mu$. This condition
leads to the appearance of the renormalization group (RG). Schematically in 
calculation of an observable, we proceed in the following way
\begin{eqnarray} \nonumber
F(k,g,m) \overset{\infty}{\longrightarrow}  
F_{\mathrm{reg}}(k,{M},g,m) \overset{M\to\infty}{\longrightarrow}  
F_{\mathrm{ren}}(k,{\mu},g,m) \overset{RG}{\longrightarrow}  
F_{\mathrm{phys}}(k,{\Lambda},m),
\end{eqnarray}
where $\Lambda$ is a dimensionful scale.
Charge (and mass) become \emph{running}, \ie energy-dependent: 
\begin{eqnarray}
g \to g\left(g,\frac{\mu'}{\mu}\right), \qquad
\beta(g) \equiv \left. \frac{\mathrm{d} g}{\mathrm{d}\ln\mu}\right|_{g_B=Const}.
\end{eqnarray}

Note that the renormalization scale $\mu$ unavoidably appears in any scheme.
Scheme and scale dependencies are reduced after including
higher and higher orders of the perturbation theory.

At this point we stop the brief introduction to Quantum field theory,
comprehensive details can be found in textbooks, 
\eg Refs.~\cite{Bogoliubov_Shirkov,Weinberg_book,Peskin_Schroeder}.

\section{Construction of the Standard Model
\label{sect:SM}}

\subsection{The Fermi model and Cabibbo--Kobayashi--Maskawa mixing matrix}

\begin{figure}
\centering\includegraphics[width=.4\linewidth]{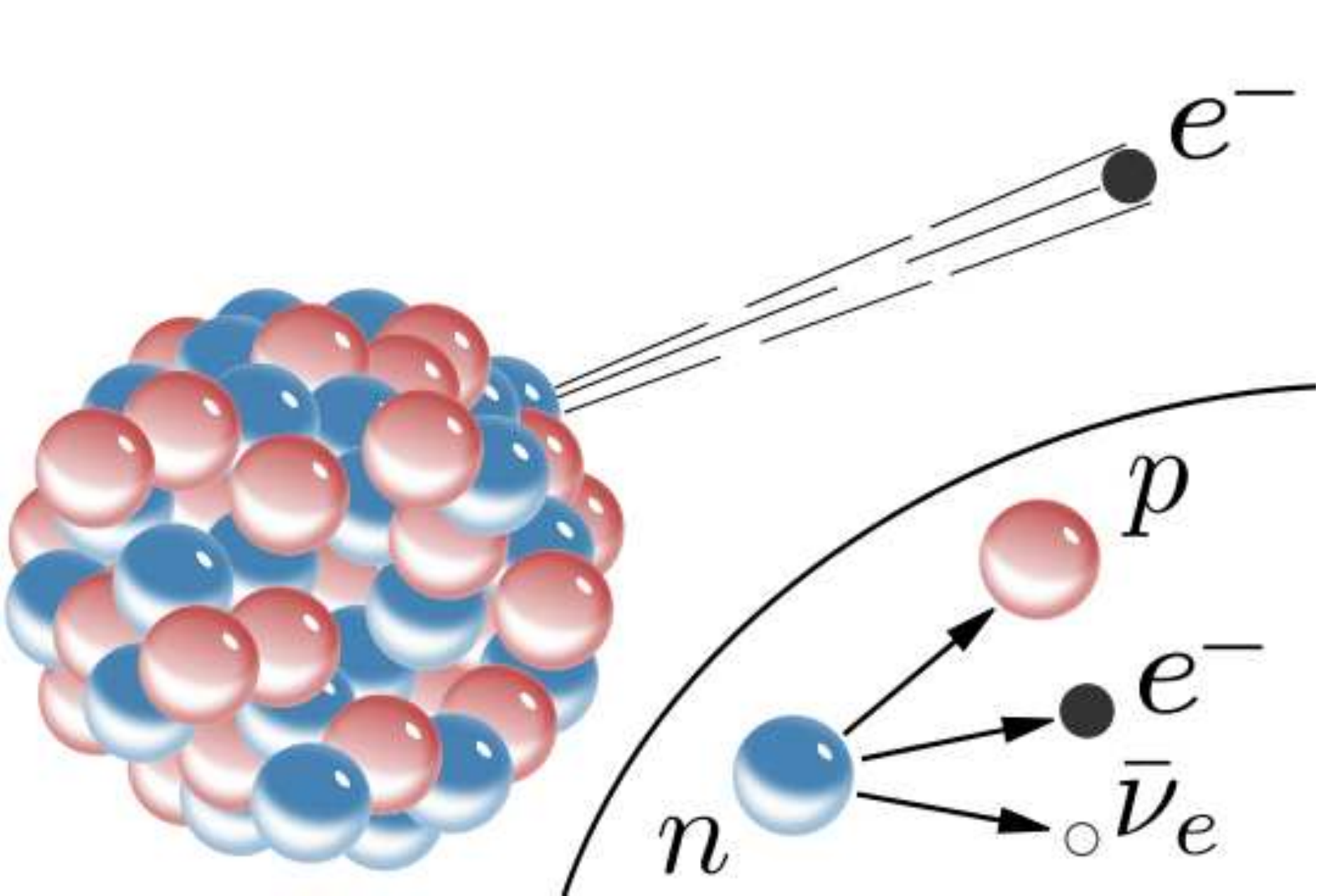}
\caption{Beta decay.}
\label{fig:beta_decay}
\end{figure}

To describe the $\beta$-decay $ n \to p + e^- + \nu_e$ in 1933, 
see \Fref{fig:beta_decay}, Enrico Fermi suggested a simple model:
\begin{eqnarray} \nonumber
\mathcal{L}_{int} = G\underbrace{\overline{\Psi}_n\gamma_\rho\Psi_p}_{J_\rho^{(N)}}\cdot
\underbrace{\overline{\Psi}_\nu\gamma_\rho\Psi_e}_{J^{(l)\dagger}_\rho}
+ h.c.
\end{eqnarray}
with interactions in the form of a product of two vector currents.
This model was inspired by QED where similar vector currents appear.

In 1957 R.~Marshak \& G.~Sudarshan; and 
R.~Feynman \& M.~Gell-Mann modified the model:
\begin{eqnarray} \label{L_Fermi}
&& \mathcal{L}_{\mathrm{Fermi}} = \frac{G_\mathrm{Fermi}}{\sqrt{2}}J_\mu J^\dagger_\mu,
\nonumber \\
&& J_\mu = \overline{\Psi}_{e}\gamma_\rho\frac{1-\gamma_5}{2}\Psi_{\nu_e}
+ \overline{\Psi}_{\mu}\gamma_\rho\frac{1-\gamma_5}{2}\Psi_{\nu_\mu}
+ (V-A)_{\mathrm{nucleons}} + h.c.
\end{eqnarray}
Explicit \emph{V-A} (Vector minus Axial--vector) form of weak interactions means 
the 100\% violation of parity. 
In fact, it appears that only left fermions
participate in weak interactions, while right fermions don't. Please remind that
massive left fermions are not states with a definite spin.
The modification of the model was required to describe differential distributions 
of $beta$-decays. Note that the CP symmetry in Lagrangian~(\ref{L_Fermi}) 
is still preserved.

The modern form of the Fermi Lagrangian includes 3 fermion generations:
\begin{eqnarray} \nonumber
\!\! \mathcal{L}_{\mathrm{Fermi}} = 
\frac{G_\mathrm{Fermi}}{\sqrt{2}} (\overline{e}_L\;\overline{\mu}_L\;\overline{\tau}_L)
\gamma_\rho\! \left(\begin{array}{c}\!\! \nu_{e,L}\!\! \\ \!\! \nu_{\mu,L}\!\! \\ 
\!\! \nu_{\tau,L}\!\! \end{array} \right) \cdot
(\overline{u}_L'\;\overline{c}_L'\;\overline{t}_L') V_u^\dagger
\gamma_\rho V_d \left(\begin{array}{c}\!\! d_{L}'\!\! \\ \!\! s_{L}'\!\! \\ \!\! b_{L}'\!\! 
\end{array} \right) + \ldots
\end{eqnarray}
Quarks $\{q'\}$ are the \emph{eigenstates} of the strong interactions, and
$\{q\}$ are the eigenstates of the weak ones.
 
Matrixes $V_{d}$ and $V_{u}$ describe quark mixing (see details in lectures 
on Flavour Physics):
\begin{eqnarray} \nonumber
\left(\begin{array}{c}\!\! d\!\! \\ \!\! s\!\! \\ \!\! b\!\! \end{array} \right)
= V_d \times \left(\begin{array}{c}\!\! d'\!\! \\ \!\! s'\!\! \\ \!\! b'\!\! \end{array} \right),\qquad
V^\dagger_u V_d \equiv {V_{\mathrm{CKM}}} = \left(\begin{array}{ccc} V_{ud} & V_{us} & V_{ub} \\
V_{cd} & V_{cs} & V_{cb} \\ V_{td} & V_{ts} & V_{tb} \end{array} \right).
\end{eqnarray}
By construction, in this model (and further in the SM) the mixing matrixes are 
unitary: ${V_i}^\dagger V_i = \mathbf{1}$. In a sense, this property just keeps the number
of quarks during the transformation to be conserved.
$V_{\mathrm{CKM}}$ contains 4 independent parameters: 3 angles and 1 phase.

\emph{ QUESTION: What is mixed by $V_{\mathrm{CKM}}$? E.g., what is mixed by the $V_{ud}$ 
element of $V_{\mathrm{CKM}}$?}

The Fermi model describes $\beta$-decays and the muon decay 
$\mu\to e+\bar{\nu}_e+\nu_\mu$ with a very high precision.
Nevertheless, there are two critical problems: \\
1. The model is non-renormalizable, remind that the dimension of the Fermi coupling constant 
$[G_{\mathrm{Fermi}}] = -2$. \\
2. Unitarity in this model is violated: consider, \eg within the Fermi model 
the total cross section of electron-neutrino scattering
\begin{eqnarray}
\sigma_{\mathrm{total}}(e\nu_e\to e\nu_e) \sim \frac{G_{\mathrm{Fermi}}^2}{\pi}s,
\qquad s = (p_e+p_{\nu_e})^2.
\end{eqnarray}
This cross section obviously growth with energy. 
Meanwhile the unitarity condition for $l^{\mathrm{th}}$ partial wave 
in the scattering theory requires that
$ \sigma_l < \frac{4\pi(2l+1)}{s} $.    
For $l=1$ we reach the \emph{unitarity limit} at
$s_0 = {2\pi\sqrt{3}}/{G_{\mathrm{Fermi}}} \approx 0.9\cdot 10^6\ \mathrm{GeV}^2 $.
So at energies above $\sim 10^3$~GeV the Fermi model is completely senseless
and somewhere below this scale another model should enter the game.

\subsection{(Electro)Weak interactions in SM}

The modern point of view is: \emph{a renormalizable QFT model which preserves unitarity
is a Yang--Mills (non-abelian) gauge model}. So we have to try to construct
an interaction Lagrangian using the principle of gauge symmetry. 

\begin{figure}
\centering\includegraphics[width=.4\linewidth]{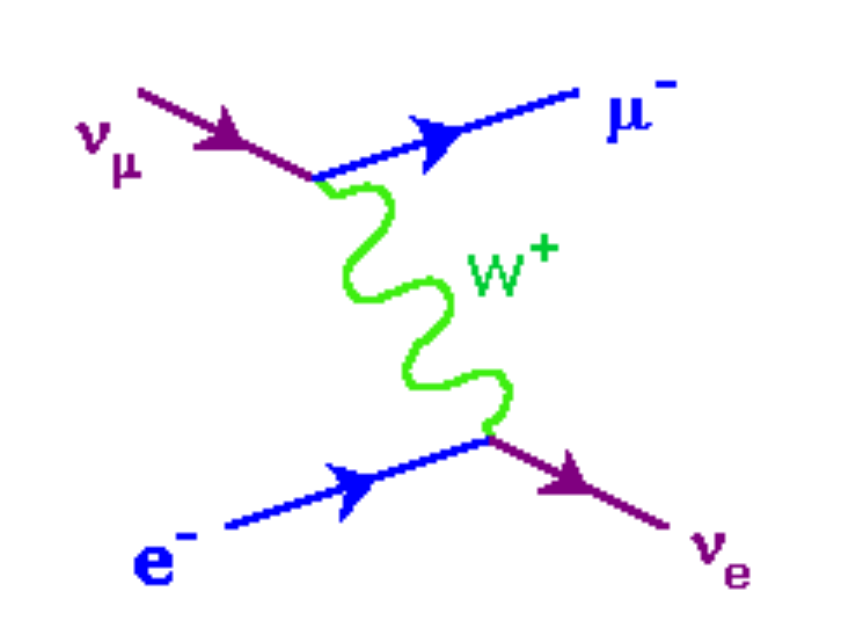}
\caption{Feynman diagram for electron-neutrino scattering with $W$ boson exchange.}
\label{fig:weak}
\end{figure}
 
Let's try to do that for description of weak interactions. 
At the first step we introduce a massive vector $W$ boson
\begin{eqnarray}
\mathcal{L}_{\mathrm{int}} = - g_w(J_\alpha W_\alpha + J_\alpha^\dagger W_\alpha^\dagger ). 
\end{eqnarray}
Then the scattering amplitude, see~\Fref{fig:weak}, takes the form
\begin{eqnarray}
T = i(2\pi)^4 g_w^2 J_\alpha \; \frac{g_{\alpha\beta}- k_\alpha k_\beta/M_W^2}{k^2-M_W^2}\; 
J_\beta^\dagger ,
\end{eqnarray}
where $k$ is the $W$ boson momentum.
If $|k| \ll M_W$ we reproduce the Fermi model with 
\begin{equation*}
\frac{G_{\mathrm{Fermi}}}{\sqrt{2}} = \frac{g_w^2}{M_W^2}. 
\end{equation*}
However such a way to introduce interactions again leads
to a non-renormalizable model. The problem appears due to the specific 
momentum dependence in the propagator of a massive vector particle, 
see Eq.~(\ref{eq:vector_prop}). Moreover, the mass term of the gauge boson
is not gauge invariant.

The \emph{minimal} way to introduce electromagnetic and weak interactions 
as gauge ones is to take the group $SU(2) \otimes U(1)$.
The abelian group $U(1)$ is the same as the one that gives conservation 
of the electric charge in QED.
Instead of the electric charge $Q$ we introduce now the \emph{hypercharge} $Y$. 
The $U(1)$ gauge symmetry provides interactions of fermions with a massless 
vector (photon-like) field $B_\mu$. 
The non-abelian group $SU(2)$ is the same as the one used for description of spinors
in Quantum mechanics. Instead of spin we use here the notion of \emph{weak isospin} $I$.
There are three massless vector Yang--Mills bosons in the adjoint representation 
of this group: $W_\mu^a$,\ $a=1,2,3$. Two of them can be electrically charged 
and the third one should be neutral. 
Introduction of the third (electro)weak boson is unavoidable,
even so that we had not have experimental evidences of weak neutral currents
at the times of the SM invention. \\
\emph{ QUESTION: Why weak interactions in the charged current (like muon and beta decays)
were discovered experimentally much earlier than the neutral current ones?} 

One can show that the model built above for gauge $SU(2) \otimes U(1)$ interactions 
of fermions and vector bosons is renormalizable and unitary. 
But this model doesn't describe the reality since all gauge bosons should be massless
because of the gauge symmetry condition. To resolve this problem we need a mechanism that
will provide masses for some vector bosons without an explicit breaking of the
gauge symmetry.

\subsection{The Brout--Englert--Higgs mechanism}

Let's consider the simple abelian $U(1)$ symmetry for interaction of a charged
scalar field $\varphi$ with a vector field $A_\mu$:
\begin{eqnarray} \nonumber
\mathcal{L} = \partial_\mu \varphi^* \partial_\mu \varphi
- V(\varphi) - \frac{1}{4}F_{\mu\nu}^2
+ i e(\varphi^* \partial_\mu \varphi - \partial_\mu \varphi^* \varphi) A_\mu
+ e^2 A_\mu A_\mu \varphi^* \varphi.
\end{eqnarray}
If $V(\varphi) \equiv V(\varphi^*\cdot \varphi)$,
$\mathcal{L}$ is invariant with respect to local $U(1)$ gauge transformations
\begin{eqnarray}
\varphi\to e^{ie\omega(x)}\varphi, \quad \varphi^*\to e^{-ie\omega(x)}\varphi^*, \quad
A_\mu \to A_\mu + \partial_\mu \omega(x). 
\end{eqnarray}

In polar coordinates $\varphi\equiv\sigma(x)e^{i\theta(x)}$ and 
$\varphi^*\equiv\sigma(x)e^{-i\theta(x)}$ and the Lagrangian takes the form
\begin{eqnarray} 
\mathcal{L} = \partial_\mu \sigma \partial_\mu \sigma
+ e^2\sigma^2\underbrace{(A_\mu-\frac{1}{e}\partial_\mu\theta)}_{\equiv B_\mu}
\underbrace{(A_\mu-\frac{1}{e}\partial_\mu\theta)}_{\equiv B_\mu}
-V(\varphi^* \varphi) - \frac{1}{4}F_{\mu\nu}^2.
\end{eqnarray}
Note that after the change of variables 
$A_\mu+\frac{1}{e}\partial_\mu\theta \to B_\mu$,
we have $F_{\mu\nu}(A) = F_{\mu\nu}(B)$ since $\theta(x)$ 
is a double differentiable function. 

We see that $\theta(x)$ is completely swallowed by the field $B_\mu(x)$.
So we made just a change of variables. But which set of variables is the 
true physical one? This question is related to the choice of variables
in which the secondary quantization should be performed.
And the answer can be given by measurements. In fact,
according to Quantum mechanics only quantum eigenstates can be observed,
so we have a reference point. Another argument can be given by a
condition on the system stability. 

\begin{figure}
\centering\includegraphics[width=.4\linewidth]{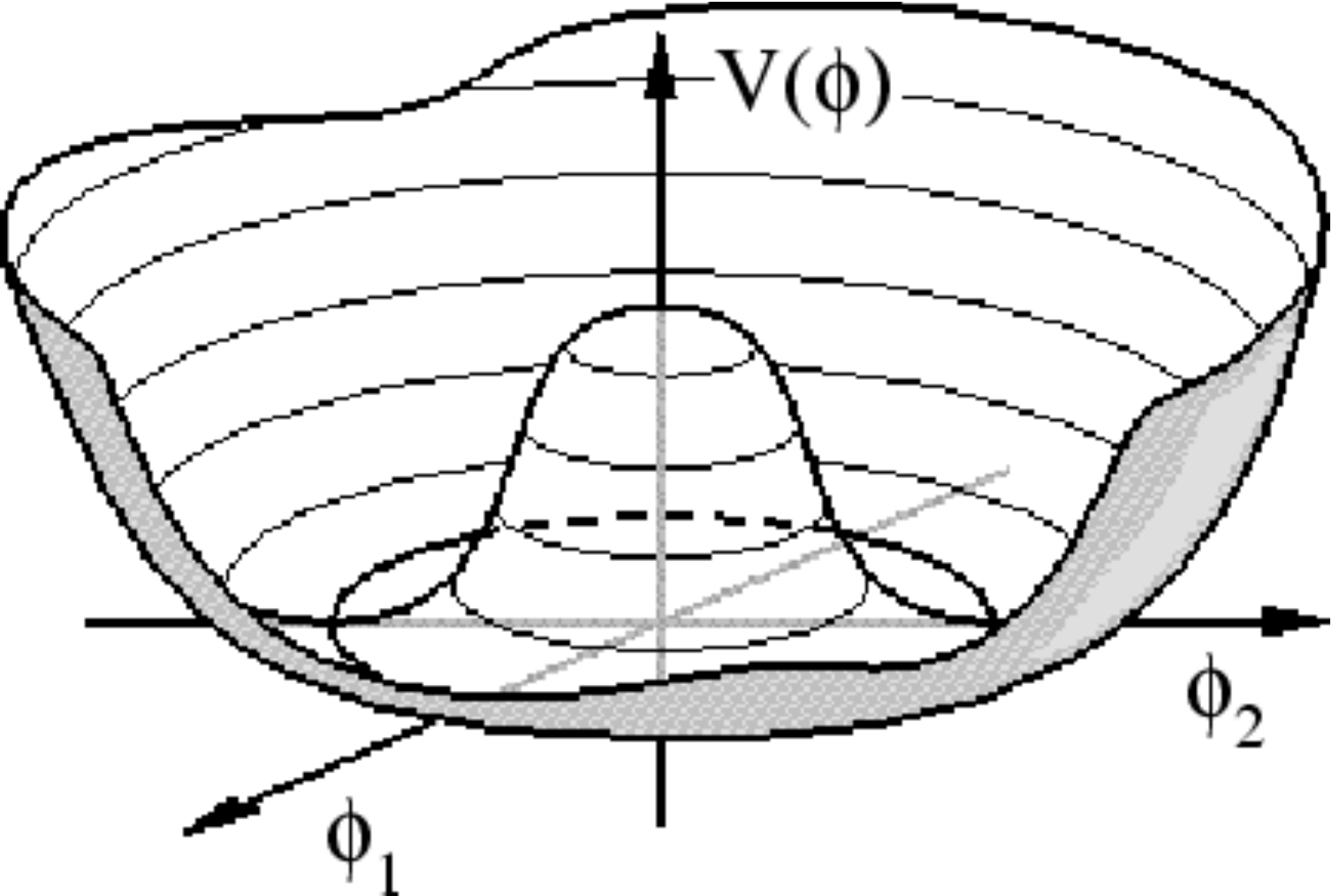}
\caption{The Higgs field potential. Picture courtesy: E.P.S. Shellard, DAMTP, Cambridge. 
From http://www.geocities.com/CapeCanaveral/ 2123/breaking.htm.}
\label{fig:Mexican}
\end{figure}
 
R.~Brout \& F.~Englert~\cite{Englert:1964et}, and P.~Higgs~\cite{Higgs:1964pj}, 
see also a brief review in the Scientific Background on 
the Nobel Prize in Physics 2013~\cite{BEH_2013},
%(following Ginzburg \& Landau) 
suggested to take the scalar potential in the form
\begin{eqnarray}
V(\varphi^*\varphi) = \lambda (\varphi^*\varphi)^2 + m^2\varphi^*\varphi.
\end{eqnarray}
For $\lambda>0$ and $m^2<0$ we get the shape of a \emph{Mexican hat}, 
see \Fref{fig:Mexican}.
We have chosen a potential for which $V(\varphi^*\varphi)=V(\sigma^2)$, 
while $\theta(x)$ corresponds to the rotational symmetry of the potential. 

By looking at the derivative of the potential
$ \frac{\mathrm{d} V(\sigma)}{\mathrm{d} \sigma} = 0$, we find two critical
points: $\sigma=0$ is the local maximum, 
and $\sigma_0 = \sqrt{-\frac{m^2}{2\lambda}}$ is the global minimum.
The stability condition suggest to shift from zero to the global minimum: 
$\sigma(x) \to h(x) + \sigma_0$. 
So we get
\begin{eqnarray} 
\mathcal{L} = \partial_\mu h \partial_\mu h
+ e^2h^2 B_\mu B_\mu + 2e^2\sigma_0 h B_\mu B_\mu  + e^2\sigma_0^2 B_\mu B_\mu 
-V(h) - \frac{1}{4}F_{\mu\nu}^2.
\end{eqnarray}
We see that field $B_\mu$ got the mass 
\begin{eqnarray}
m_B^2=2e^2\sigma_0^2 = - \frac{e^2 m^2}{\lambda} > 0.
\end{eqnarray}
So, we generated a mass term for the vector field without putting it into
the Lagrangian by hand. That is the core of the Brout--Englert--Higgs
mechanism.

The quantity $\sigma_0 \equiv v $ is the \emph{vacuum expectation value} 
(vev) of $\sigma(x)$,
\begin{eqnarray}
v \equiv \langle 0|\sigma|0\rangle, \qquad 
v = \frac{1}{V_0} \int_{V_0} \mathrm{d}^3 x\; \sigma(x).
\end{eqnarray}

Look now at the potential (keep in mind $m^2 = - 2\lambda v^2$)
\begin{eqnarray} \nonumber
V(h)\!\! &=&\!\!  \lambda(h+v)^4+m^2(h+v)^2
\\ \nonumber
\!\! &=&\!\!  \lambda h^4 + 4\lambda vh^3 
+ h^2\underbrace{(6\lambda v^2+ m^2)}_{2m_h^2=4\lambda v^2} 
+ h\underbrace{(4\lambda v^3 + 2m^2v)}_{=0}
+ \lambda v^4 + m^2v^2. 
\end{eqnarray}
So, the scalar field $h$ has a normal (not tachyon-like) mass term, $m_h^2>0$.
One can see that the initial tachyons $\varphi$ are not physically observable,
since they are not pure states in the basis of the secondary quantized system 
of fields.

It is worth to note that even so that the field content of the Lagrangian is
changed, but the number of degrees of freedom is conserved. In fact initially
we had two components of the scalar field and two components of the massless
vector field, and after the change of variables we have a single scalar field
plus a massive vector field with 3 independent components: $2+2=1+3$.

The field $\theta(x)$ is a \emph{Nambu--Goldstone boson} (a goldstone). 
It is massless, $m_\theta = 0$,
and corresponds to effortless rotations around the vertical symmetry axis of the potential.
In general, the Goldstone theorem claims that in a model with spontaneous breaking of
a continuous global symmetry $G_n$ (remind the first Noether theorem) there exist
as many massless modes, as there are group generators which do not preserve the
vacuum invariance.

The constant term $\lambda v^4 + m^2v^2$ obviously doesn't affect
equations of motion in QFT, but it contributes to the Universe energy density
(too much, actually).
That makes a problem for Cosmology. Formally, one can make a shift
of the initial Lagrangian just by this term and avoid the problem
at the present time of the Universe evolution. 

One has to keep in mind that the term ``spontaneous breaking of the gauge 
symmetry'' is just a common notation, while in fact a (local) gauge symmetry
can not be broken spontaneously as proved by S.~Elitzur~\cite{Elitzur:1975im}.
A detailed discussion can be found in~\cite{Chernodub:2008rz}, 
see also~\cite{Wilczek:2012sb}. 

Now let us return to the case of the Standard Model.
To generate masses for 3 vector bosons we need at least
3 goldstones. The minimal possibility is to introduce
one complex scalar doublet field:
\begin{eqnarray} 
&& \Phi \equiv \left(\begin{array}{c} \Phi_1 \\ \Phi_2 \end{array}\right),\qquad
\Phi^\dagger = (\Phi^*_1\ \ \Phi^*_2). 
\end{eqnarray}
Then the following Lagrangian is $SU(2)\otimes U(1)$ invariant  
\begin{eqnarray} \nonumber
&& \mathcal{L} = (D_\mu\Phi)^\dagger(D_\mu\Phi) - m^2\Phi^\dagger\Phi
- \lambda(\Phi^\dagger\Phi)^2 - \frac{1}{4}W_{\mu\nu}^aW_{\mu\nu}^a
- \frac{1}{4}B_{\mu\nu}B_{\mu\nu},
\\ \nonumber
&& B_{\mu\nu} \equiv \partial_{\mu}B_\nu - \partial_{\nu}B_\mu, \qquad
W_{\mu\nu}^a \equiv \partial_{\mu}W_\nu^a 
- \partial_{\nu}W_\mu^a + g\varepsilon^{abc}W_\mu^b W_\nu^c, 
\\ 
&& D_\mu\Phi \equiv \partial_\mu\Phi + igW_{\mu}^a\frac{\tau^a}{2}\Phi
+ \frac{i}{2}g'B_\mu\Phi. %, \qquad D_\mu\Phi^\dagger = \ldots
\end{eqnarray}
Again for $m^2<0$ there is a non-trivial minimum of the Higgs
potential and a non-zero vev of a component: $\langle 0|\Phi_2|0\rangle=\eta/\sqrt{2}$.
In accord with the Goldstone theorem, three massless bosons appear. The
global $SU(2)\times SU(2)$ symmetry of the Higgs sector is reduced to the 
\emph{custodial} $SU(2)$ symmetry.

\subsection{Electroweak bosons}

The gauge bosons of the $SU(2)\otimes U(1)$ group can be represented as
\begin{eqnarray} 
W_\mu^+ = \frac{W_\mu^1+iW_\mu^2}{\sqrt{2}},\qquad
W_\mu^- = \frac{W_\mu^1-iW_\mu^2}{\sqrt{2}},\qquad
W_\mu^0 = W_\mu^3, \qquad B_\mu.
\end{eqnarray}  
$W^0_\mu$ and $B_\mu$ are both neutral and have the same quantum numbers,
so they can mix. In a quantum world, ``can'' means ``do'':
\begin{eqnarray} \nonumber
&& W_\mu^0 =\ \cos\theta_w\, Z_\mu + \sin\theta_w\, A_\mu, 
\\ 
&& B_\mu = - \sin\theta_w\, Z_\mu + \cos\theta_w\, A_\mu, 
\end{eqnarray} 
where $\theta_w$ is the \emph{weak mixing angle}, 
introduced first by S.~Glashow, 
$\theta_w$ is known also the Weinberg angle.
Remind that we have to choose variables which correspond to
observables. Vector bosons $Z_\mu$ and $A_\mu$ are linear combinations
of the primary fields $W_\mu^0$ and $B_\mu$.

It is interesting to note that Sheldon Glashow, Abdus Salam, and Steven Weinberg 
have got the Nobel Prize in 1979, \emph{before} the discovery of $Z$ and $W$ bosons in 1983,
and even much longer before the discovery of the Higgs boson.
So the Standard Model had been distinguished before experimental confirmation
of its key components.   

Look now at the scalar fields:
\begin{eqnarray} \nonumber
\Phi \equiv \frac{1}{\sqrt{2}}\left(\begin{array}{c}\Psi_2(x)+ i\Psi_1(x) \\
\eta + \sigma(x) + i\xi(x) \end{array}\right), \qquad
\Phi^\dagger = \ldots
\end{eqnarray}  
Fields $\Psi_{1,2}$ and $\xi$ become massless Goldstone bosons.
We \emph{hide} them into the vector fields:
\begin{eqnarray} \nonumber
&& W_\mu^i \to W_\mu^i + \frac{2}{g\eta}\partial_\mu\Psi_i\ \ \Rightarrow \ \ 
M_W=\frac{g\eta}{2},
\nonumber \\
&& Z_\mu = \frac{g}{\sqrt{g^2+{g'}^2}}W_\mu^0 - \frac{g'}{\sqrt{g^2+{g'}^2}}B_\mu
- \frac{2}{\eta\sqrt{g^2+{g'}^2}} \partial_\mu\xi 
\ \  \Rightarrow \ \ 
M_Z=\frac{\eta\sqrt{g^2+{g'}^2}}{2}.
\end{eqnarray}
The photon field appears massless by construction.
Looking at the mixing we get 
\begin{eqnarray}\nonumber
\cos\theta_w =  \frac{g}{\sqrt{g^2+{g'}^2}} = \frac{M_W}{M_Z}. 
\end{eqnarray}

The non-abelian tensor
\begin{eqnarray} \nonumber
W_{\mu\nu}^a \equiv \partial_{\mu}W_\nu^a 
- \partial_{\nu}W_\mu^a + g\varepsilon^{abc}W_\mu^b W_\nu^c 
\end{eqnarray}
leads to triple and quartic self-interactions of the \emph{primary}
$W^a_\mu$ bosons, since
\begin{eqnarray}
\mathcal{L} = - \frac{1}{4}W_{\mu\nu}^aW_{\mu\nu}^a + \ldots 
\end{eqnarray}
Fields $B_\mu$ and $W^a_\mu$ were not interacting between each other.
But after the spontaneous breaking of the global symmetry in the Higgs sector, 
and the consequent change of the basis $\{W_\mu^0,B_\mu\}\to\{Z_\mu,A_\mu\}$, 
we get interactions of charged $W^\pm_\mu$ bosons with photons. And the charge
of the physical $W$ bosons is well known from the condition of charge conservation
applied to $beta$-decays. That allows to fix the relation between the constants:
\begin{eqnarray}
e = \frac{gg'}{\sqrt{g^2+{g'}^2}} = g\sin\theta_w.
\end{eqnarray}
We see that the very construction of the SM requires phenomenological
input. So on the way of the SM building, not everything comes 
out automatically from symmetry principles \etc

\subsection{EW interactions of fermions}

We have chosen the $SU(2)\otimes U(1)$ symmetry group. 
To account for parity violation in weak decays, we assume different behavior 
of left and right fermions under $SU(2)_L$ transformations:
\begin{eqnarray} \nonumber
&& \mathrm{left\ doublets} \quad 
\left(\begin{array}{c} \nu_e \\ e \end{array}\right)_L,\ \ 
\left(\begin{array}{c} u \\ d \end{array}\right)_L\ \  + 2\ \mathrm{other\ generations},
\\ \nonumber 
&& \mathrm{right\ singlets} \quad 
e_R,\ u_R,\ d_R,\ (\nu_{e,R})\ \  + 2\ \mathrm{other\ generations}.
\end{eqnarray}
To preserve the gauge invariance, the fermion Lagrangian is constructed 
with the help of covariant derivatives:
\begin{eqnarray} \nonumber
&& \mathcal{L}(\Psi) =\sum_{\Psi_i}\biggl[ \frac{i}{2}\biggl(\overline{\Psi}_L
\gamma_\alpha D_\alpha \Psi_L - D_\alpha\overline{\Psi}_L\gamma_\alpha \Psi_L \biggr)
% \\ \nonumber && \qquad
+ \frac{i}{2}\biggl(\overline{\Psi}_R
\gamma_\alpha D_\alpha \Psi_R - D_\alpha\overline{\Psi}_R\gamma_\alpha \Psi_R \biggr)
\biggr],
\\ \nonumber 
&& D_\alpha \Psi_L \equiv \partial_\alpha\Psi_L + \frac{ig\tau^b}{2}W_\alpha^b\Psi_L
- i{g_1}B_\alpha\Psi_L,
\qquad
D_\alpha \Psi_R \equiv \partial_\alpha\Psi_L - i{g_2}B_\alpha\Psi_L.
\end{eqnarray}
All interactions of the SM fermions with electroweak vector bosons are here.
But coupling constants $g_{1,2}$ still have to be fixed and related to observables.

Fermions have weak isospins and hypercharges $(I,Y)$:
\begin{eqnarray} 
\Psi_L: \quad \biggl(\frac{1}{2}, \ \ - \frac{2g_1}{g'} \biggr), 
\qquad
\Psi_R: \quad \biggl(0, \ \ - \frac{2g_2}{g'} \biggr). 
\end{eqnarray}
Looking at interactions of left and right electrons with $A_\mu$ in
$\mathcal{L}(\Psi)$ we fix their hypercharges:
\begin{eqnarray} 
e_L: \quad \biggl(-\frac{1}{2}, \ \ - 1 \biggr), 
\qquad
e_R: \quad \biggl(0, \ \ - 2 \biggr). 
\end{eqnarray}
The \emph{Gell-Mann--Nishijima formula} works for all fermions:
\begin{eqnarray}
Q = I_3 + \frac{Y}{2},
\end{eqnarray}
where $Q$ is the electric charge of the given fermion, 
$I_3$ is its weak isospin projection, 
and $Y$ is its hypercharge.

Interactions of leptons with $W^\pm$ and $Z$ bosons come out in the form
\begin{eqnarray} \nonumber
&& \mathcal{L}_I = - \frac{g}{\sqrt{2}}\bar{e}_L\gamma_\mu\nu_{e,L}W_\mu^- + h.c.
- \frac{gZ_\mu}{2\cos\theta_w}\biggl[ \bar{\nu}_{e,L}\gamma_\mu \nu_{e,L}
\\ \nonumber && \quad 
+ \bar{e}\gamma_\mu\biggl( - (1-2\sin^2\theta_w)\frac{1-\gamma_5}{2}
+ 2\sin^2\theta_w\frac{1+\gamma_5}{2}\biggr)e
\biggr] 
\\ \nonumber && \Rightarrow 
g_w = \frac{g}{2\sqrt{2}}, \quad M_W^2 = \frac{g^2\sqrt{2}}{8G_{\mathrm{Fermi}}}
= \frac{e^2\sqrt{2}}{8G_{\mathrm{Fermi}}\sin^2\theta_w}
= \frac{\pi\alpha}{\sqrt{2}G_{\mathrm{Fermi}}\sin^2\theta_w}.
\end{eqnarray}
That gives $M_W = \frac{38.5}{\sin\theta_w}$~GeV, 
remind $M_Z=\frac{M_W}{\cos\theta_w}$. 

We can see that the Higgs boson vev is directly related to the Fermi coupling 
constant: 
\begin{eqnarray}
v = (\sqrt{2}G_{\mathrm{Fermi}})^{-1/2}\approx 246.22\ \mathrm{GeV}.
\end{eqnarray}
So this quantity had been known with a high precision long before
the discovery of the Higgs boson and the experimental measurement of its mass.

\emph{ QUESTION: Why neutral weak currents in the SM do not change flavour
(at the tree level)?}

\subsection{Self-interactions of EW bosons and Faddeev--Popov ghosts}

\begin{figure}
\centering\includegraphics[width=.8\linewidth]{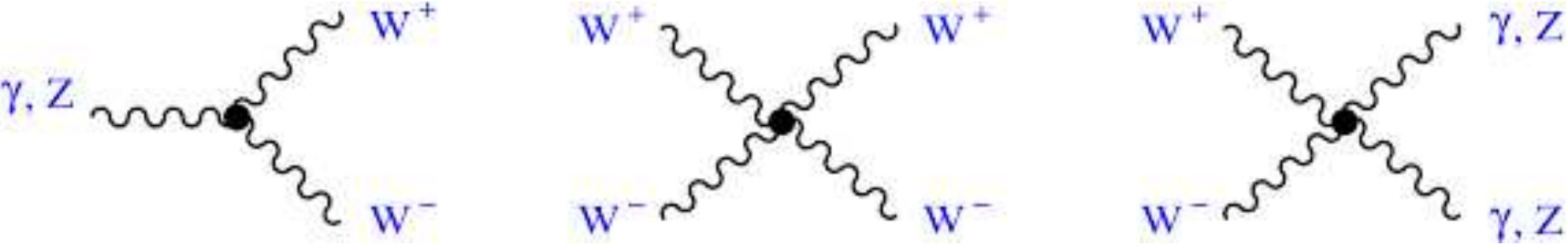}
\caption{Vertexes of EW boson self-interactions.}
\label{EW_self}
\end{figure}

Because of the non-abelian $SU(2)_L$ group structure and mixing of the neutral vector
bosons, we have a rather reach structure of EW boson self-interactions, 
see~\Fref{EW_self}. The corresponding contributions to the SM Lagrangian
look as follows:
\begin{eqnarray} \nonumber
&&\!\!\!\! \mathcal{L}_3 \sim ie\frac{\cos\theta_w}{\sin\theta_w}
\bigg[ (\partial_\mu W^-_\nu - \partial_\nu W^-_\mu)W_\mu^+ Z_\nu
- (\partial_\mu W^+_\nu - \partial_\nu W^+_\mu)W_\mu^- Z_\nu
\\ \nonumber && 
+ W_\mu^-W_\nu^+(\partial_\mu Z_\nu - \partial_\nu Z_\mu) \biggr]
\\ \nonumber 
&&\!\!\!\! \mathcal{L}_4 \sim - \frac{e^2}{2\sin^2\theta_w}\biggl[
(W_\mu^+ W_\mu^-)^2 - W_\mu^+W_\mu^+W_\nu^-W_\nu^- \biggr],
\\ \nonumber && 
- \frac{e^2\cos^2\theta_w}{\sin^2\theta_w}\biggl[
W_\mu^+W_\mu^-Z_\nu Z_\nu 
- W_\mu^+Z_\mu W_\mu^- Z_\nu \biggr]
\\ \nonumber && 
- \frac{e^2\cos^2\theta_w}{\sin^2\theta_w}\biggl[
2W_\mu^+W_\mu^-Z_\nu A_\nu - W_\mu^+Z_\mu W_\mu^- A_\nu 
- W_\mu^+A_\mu W_\mu^- Z_\nu \biggr]
\\ \nonumber && 
- e^2\biggl[ W_\mu^+W_\mu^-A_\nu A_\nu - W_\mu^+A_\mu W_\mu^- A_\nu \biggr].
\end{eqnarray}

As we discussed earlier, an accurate treatment of non-abelian gauge symmetries
leads to introduction of Faddeev--Popov ghosts.
For the $SU(2)$ case we obtain 3 ghosts: $c_a(x),\ \ a=1,2,3$, 
\begin{eqnarray} \nonumber
&& c_1 = \frac{X^++X^-}{\sqrt{2}}, \quad
c_2 = \frac{X^+-X^-}{\sqrt{2}}, \quad
c_3 = Y_Z\cos\theta_w - Y_A\sin\theta_w ,
\\ \nonumber
&& \mathcal{L}_{gh} = \underbrace{\partial_\mu \bar{c}_i 
(\partial_\mu c_i - g\varepsilon_{ijk}c_jW_\mu^k)}_{\mathrm{kinetic\ +\ int.\ with}\ W^a}
+ \underbrace{\mathrm{int.\ with}\ \Phi}_{M_{gh},\ \mathrm{int.\ with}\ H}.   
\end{eqnarray}
Propagators of the ghost fields read
\begin{eqnarray} \nonumber
D_{Y_\gamma}(k) = \frac{i}{k^2+i0}, \qquad
D_{Y_Z}(k) = \frac{i}{k^2-\xi_ZM_Z^2+i0}, \qquad
D_{X}(k) = \frac{i}{k^2-\xi_WM_W^2+i0},
\end{eqnarray}
where $\xi_i$ are the gauge parameters.
Note that masses of the ghosts $Y_\gamma$, $Y_Z$, and $X^\pm$ coincide
with the ones of photon, $Z$, and $W^\pm$, respectively. That is important
for \emph{gauge invariance} of total amplitudes. 
The ghosts appear only in propagators, but not in the final or initial 
asymptotic states.

\subsection{Generation of fermion masses}

We observe massive fermions, but the $SU(2)_L$ gauge symmetry
forbids fermion mass terms, since
\begin{eqnarray}
m\overline{\Psi}\Psi = m\biggl(\overline{\Psi}\frac{1+\gamma_5}{2}
+\overline{\Psi}\frac{1-\gamma_5}{2}\biggr)
\biggl(\frac{1+\gamma_5}{2}\Psi+\frac{1-\gamma_5}{2}\Psi\biggr)=
m(\overline{\Psi}_L\Psi_R + \overline{\Psi}_R\Psi_L ) 
\end{eqnarray}
while $\Psi_L$ and $\Psi_R$ are transformed in different ways under $SU(2)_L$.
The SM solution is to introduce Yukawa interactions of fermions with the primary
Higgs boson doublet field:
\begin{eqnarray} \nonumber
\mathcal{L}_Y &=& - y_d (\bar{u}_L \bar{d}_L)
\left(\begin{array}{c} \phi^+ \\ \phi^0 \end{array}\right) d_R
- y_u (\bar{u}_L \bar{d}_L)
\left(\begin{array}{c} \phi^{0*} \\ -\phi^- \end{array}\right) u_R
\\ \nonumber
&-& y_l (\bar{\nu}_L \bar{l}_L)
\left(\begin{array}{c} \phi^+ \\ \phi^0 \end{array}\right) l_R
- y_\nu (\bar{\nu}_L \bar{l}_L)
\left(\begin{array}{c} \phi^{0*} \\ -\phi^- \end{array}\right) \nu_R
+ h.c.
\end{eqnarray}
The form of this Lagrangian is fixed by the condition of the
$SU(2)_L$ gauge invariance.
It is worth to note that neutrino masses can be generated exactly
in the same way as the up quark ones. Of course, that requires introduction 
of additional Yukawa constants $y_\nu$. 
The Pontecorvo--Maki--Nakagawa--Sakata (PMNS) mixing matrix for (Dirac) 
neutrinos can be embedded in the SM.

\emph{ QUESTION: Why do we need ``$h.c.$'' in $\mathcal{L}_Y$?} 

Spontaneous breaking of the global symmetry in the Higgs sector
provides mass terms for fermions and Yukawa interactions 
of fermions with the Higgs boson: 
\begin{eqnarray} \nonumber
\mathcal{L}_Y = - \frac{v+H}{\sqrt{2}}\left[
y_d\bar{d}d + y_u\bar{u}u + y_l\bar{l}l + y_\nu\bar{\nu}\nu \right]
\quad \Rightarrow \quad
m_f = \frac{y_f}{\sqrt{2}}v.
\end{eqnarray}
By construction, the coupling of the Higgs boson to a fermion is proportional 
to its mass $m_f$.
It is interesting to note that the top quark Yukawa coupling is very close to 1.
And there is a very strong hierarchy of fermion masses: 
\begin{eqnarray} \nonumber
y_t\approx 0.99\ \ \gg\ \  y_e\approx 3\cdot 10^{-6} \ \ \gg \ \ y_{\nu}\approx ?
\end{eqnarray}
The question mark in the last case is given not only because we do not know
neutrino masses, but also since we are not sure the they are generated
by the same mechanism.

Quarks can mix and Yukawa interactions are not necessarily diagonal 
neither in the basis of weak interaction eigenstates, nor
in the basis of the strong ones. 
In the eigenstate basis of a given interaction for the case of three generations,
the Yukawa coupling constants are $3\times 3$ matrixes: 
\begin{eqnarray} \nonumber
\mathcal{L}_Y &=& - \sum\limits_{j,k=1}^3 \left\{ (\bar{u}_{jL} \bar{d}_{jL})
\left[ \left(\begin{array}{c}\!\! \phi^+\!\! \\\!\! \phi^0\!\! \end{array}\right) 
y_{jk}^{(d)} d_{kR}
+ \left(\begin{array}{c}\!\! \phi^{0*}\!\! \\ \!\! -\phi^-\!\! \end{array}\right) 
y_{jk}^{(u)}u_{kR}  \right] \right.
\\ \nonumber 
&+& \left. (\bar{\nu}_{jL} \bar{l}_{jL})
\left[\left(\begin{array}{c}\!\! \phi^+\!\! \\ \!\! \phi^0\!\! \end{array}\right) 
y_{jk}^{(l)}l_{kR}\
+  \left(\begin{array}{c}\!\! \phi^{0*}\!\! \\ \!\! -\phi^-\!\! \end{array}\right) 
y^{(\nu)}_{jk}\nu_{kR} 
\right] \right\}  + h.c.
\end{eqnarray}
where indexes $j$ and $k$ mark the generation number.

Charged lepton mixing is formally allowed in the SM, but not (yet) observed 
experimentally. Searches for lepton flavour violating processes, 
like the $\mu\to e\gamma$ decay, are being performed.

\subsection{Short form of the SM Lagrangian}

At CERN one can buy souvenirs with the Standard Model
Lagrangian represented in a very short compressed form:
\begin{eqnarray} \label{SM_lag_short}
\mathcal{L}_{\mathrm{SM}} &=& - \frac{1}{4}F_{\mu\nu}F^{\mu\nu}
\nonumber \\
&+& i\bar{\Psi} \not\!\!{D} \Psi + h.c.
\nonumber \\
&+& \Psi_i y_{ij} \Psi_j \Phi + h.c.
\nonumber \\
&+& |D_\mu\Phi|^2 - V(\Phi).
\end{eqnarray}
We can understand now the meaning of each term. First of all, we see that
the Lagrangian is given in the initial form before the spontaneous symmetry breaking.
Summation over $SU(3)_C$, $SU(2)_L$, and $U(1)_Y$ gauge groups is implicitly 
assumed in the first term. 
The second line represents the kinetic terms and gauge interactions 
of fermions provided by the covariant derivative(s).
The third line is the Yukawa interaction of fermions with the primary
scalar doublet field.
And the fourth line represents the kinetic and potential terms of the scalar field. 

\emph{ EXERCISE: Find two 'misprints' in the Lagrangian~(\ref{SM_lag_short}) 
which break the commonly accepted QFT notation discussed in Sect.~\ref{sec:QFT}}.

\subsection{Axial anomaly}

\begin{figure}
%\centering\includegraphics[width=.8\linewidth]{anomaly.pdf}
\centering\includegraphics[width=.3\linewidth]{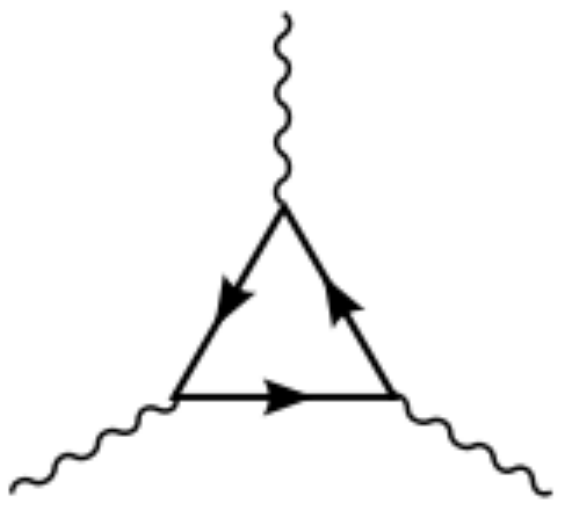}
\caption{Triangular anomaly diagram.}
\label{fig:anomaly}
\end{figure}

There are \emph{axial-vector} currents in the SM:
\begin{eqnarray}
J_\mu^A = \overline{\Psi}\gamma_\mu\gamma_5 \Psi. 
\end{eqnarray}
In the case of massless fermions, the unbroken global symmetry 
(via the Noether theorem) leads to conservation
of these currents: $\partial_\mu J_\mu=0$. 
For massive fermions $\partial_\mu J_\mu^A=2im\overline{\Psi}\gamma_5\Psi$.
But \emph{one-loop corrections}, see~\Fref{fig:anomaly}, give
\begin{eqnarray}
\partial_\mu J_\mu^A=2im\overline{\Psi}\gamma_5\Psi
+ \frac{\alpha}{2\pi} F_{\mu\nu}\tilde{F}_{\mu\nu}, \qquad
\tilde{F}_{\mu\nu} \equiv \frac{1}{2}\varepsilon_{\mu\nu\alpha\beta}F_{\alpha\beta}.
\end{eqnarray}
That fact is known as the \emph{axial} or \emph{chiral} or \emph{triangular} 
Adler--Bell--Jackiw anomaly, see~\cite{Weinberg_book} for details.
So at the quantum level the classical symmetry is lost. 
That is a real problem for the theory. In simple words, such a symmetry breaking 
makes the classical and quantum levels of the theory being inconsistent
to each other. Moreover, the resulting quantum theory looses unitarity.

But in the SM the axial anomalies apparently cancel out. This can be seen
for all possible combination of external gauge bosons: \\
1) $(W\, W\, W)$ and $(W\, B\, B)$ --- automatically since left leptons and quarks 
are doublets; \\[.2cm]
2) $(B\, W\, W)$ --- since {$Q_e+2Q_u+Q_d=0$};  \\[.2cm]
3) $(B\, B\, B)$ --- since {$Q_e=-1,\ Q_\nu=0,\ Q_u=\frac{2}{3},\ Q_d=-\frac{1}{3}$};  \\[.2cm]
4) $(B\, g\, g)$ ---  automatically ($g=$ gluon); \\[.2cm]
5) $(B\, gr\, gr)$ --- the same as '3)' ($gr=$ graviton). \\
Here $B$ and $W$ are the primary $U(1)$ and $SU(2)_L$ gauge bosons.
Note that anomalies cancel out in each generation separately.  
It is interesting to note that condition '2)' means that the hydrogen atom is neutral.

It is very important that the axial anomalies cancel out in the \emph{complete SM}: 
with the $SU(3)_C\otimes SU(2)_L\otimes U(1)_Y$ gauge symmetries. So there is a
nontrivial connection between the QCD and EW sectors of the model.

\emph{ QUESTION: Where is $\gamma_5$ in the $(B\, B\, B)$ case?}

\subsection{Parameters and interactions in the SM}

The SM has quite a lot of parameters. 
We do not know (yet) where do they come from and have to define their
values from observations. 
Let us fist count the number of independent free parameters in the SM. 
It is convenient to perform this exercise by looking at the initial form
of the SM Lagrangian before the change of variables invoked by the
spontaneous symmetry breaking. 
So, we have:
\begin{itemize}
    \item 3 gauge charges $(g_1,\ g_2,\ g_s)$;
    \item 2 parameters in the Higgs potential;
    \item 9 Yukawa couplings for charged fermions;
    \item 4 parameters in the CKM matrix.
\end{itemize}
It makes in total 18 free parameters for the \emph{canonical} Standard Model.
Sometimes, we add also as a free parameter $\theta_{\mathrm{CP}}$ which is
responsible for CP symmetry violation in the QCD sector. But at the present time 
this parameter is determined experimentally to be consistent with zero, so
we can drop it for the time being.
Moreover, we can include neutrino masses and mixing, as described above.
That would give in addition 4 (or 6 for the Majorana case) parameters in the
PMNS matrix and 3 more Yukawa couplings. 

\emph{ QUESTION: How many independent dimensionful parameters is there in the SM?}

Most likely that many of the listed parameters are not true independent ones.
There should be some hidden symmetries and relations. Those certainly go beyond
the SM. In spite of a large number of parameters the SM is distinguished 
between many other models by its minimality and predictive power. For example,
the supersymmetric extension of the SM formally has more than one hundred free
parameters, and for this reason it is not able to provide unambiguous predictions 
for concrete observables.   

Let us now count the interactions in the SM. Obviously, we should do that
in accord with the QFT rules. The key point is to exploit symmetries, 
first of all the gauge ones.
But looking at the Lagrangian it might be
not clear what actually should be counted:\\
--- number of \emph{different vertexes} in Feynman rules?\\
--- number of particle which \emph{mediate} interactions?\\
--- number of \emph{coupling constants}?

Our choice here is to count coupling constants. In fact that will
automatically help us to avoid double coupling of the same interactions.
This way how to count interactions is dictated by the QFT rules.
So we have: 
\begin{itemize}
    \item  3 gauge charges $(g_1,\ g_2,\ g_s)$;
    \item  1 self-coupling $\lambda$ in the Higgs potential;
    \item  9 Yukawa couplings for charged fermions. 
\end{itemize} 
If required we can add 3 Yukawa couplings for neutrinos.
We see that the SM contains 5 types of interactions:
3 gauge ones, the self-interaction of scalar bosons, and the Yukawa
interactions of the scalar bosons with fermions. 
Note also that even we like some interactions \eg the gauge ones, 
in the SM more than others, we can not say that any of them is more 
fundamental than others just since they all are in the same Lagrangian.

\subsection{The naturalness problem in the SM
\label{sect:natural}}

The most serious and actually the only one real theoretical problem 
of the SM is the \emph{naturalness problem} known also as \emph{fine-tuning} 
or \emph{hierarchy} one. 
Note that all but one masses in the SM are generated due to the spontaneous
symmetry breaking in the Higgs sector. While the scalar boson mass itself has been 
introduced \emph{by hands} (of Peter Higgs {\it et al.}) from the beginning.
The tachyon mass term breaks the scale invariance 
(the conformal symmetry) \emph{explicitly}.

So the running of all but one masses is suppressed by the classical symmetries.
As the result, all other masses run with energy only logarithmically, but the Higgs
mass gets quadratically divergent radiative corrections. 
In the one-loop approximation we get
\begin{eqnarray} \nonumber
M_H^2 = (M_H^0)^2 + \frac{3 \Lambda^2}{8\pi^2v^2}\biggl[M_H^2 
+ 2M_W^2 + M_Z^2\ - \ 4m_t^2 \biggr],
\end{eqnarray}
where $\Lambda$ is a formal UV cut-off. At the same time $\Lambda$ can
be the energy scale of a new physics which is coupled to the EW one.
In particular $\Lambda$ can be even the Planck mass scale.
On the other hand, it is unnatural to have $\Lambda \gg M_H$. 
The most natural option would be $\Lambda \sim M_H$ \eg everything 
is defined by the EW scale. But that is not the case of the SM\ldots
There are two general ways to solve the problem: \\
--- either to exploit some (super)symmetry to cancel out the huge terms; \\
--- or to introduce some new physics at a scale not very far from the
electroweak one, \ie making $\Lambda$ being not large. 
One can find in the literature quite a lot of models for both options.
But the experimental data coming from modern accelerators and rare decay
studies disfavor most of scenarios of new physics with scales up to about 1~TeV
and even higher. 
Moreover, it was shown that the measured value of the Higgs boson mass makes the SM 
being self-consistent up to very high energies even up to the Planck mass 
scale~\cite{Bednyakov:2015sca}.
Direct and indirect experimental searches push up and up possible energy scale 
of new physical phenomena. 
So the naturalness problem becomes nowadays more and more prominent. 
And the question, why the top quark mass, the Higgs boson mass and
and vacuum expectation value $v$ are of the same order becomes more and 
more intriguing. In a sense, the problem is not about how to deal with
divergent radiative corrections, but how to understand the very origin 
of the EW energy scale.  

After the discovery of the Higgs boson and the measurement of its mass,
we found some remarkable empirical relation between parameters of the SM.
In particular the equality
\begin{eqnarray} \label{puzzle1}
v = \sqrt{M_H^2 + M_W^2 + M_Z^2 + m_t^2} 
\end{eqnarray}
holds within the experimental errors: $246.22 = 246\pm 1$~GeV.
Obviously, there should be some tight clear relation between 
the top quark mass and the Higgs boson one (or the EW scale in general).
The present version of the SM does not explain this puzzle.

\emph{ EXERCISE: Divide both sides of Eq.~(\ref{puzzle1}) by $v$
and find a relation between coupling constants.}

Another interesting relation also involves the Higgs boson and the top quark:
\begin{eqnarray}
2\frac{m_h^2}{m_t^2} = 1.05 \approx 1  \approx 2\frac{m_t^2}{v^2}\equiv y_t^2 = 0.99.
\end{eqnarray}
It might be that these relations are of a pure numerological nature, but they
certainly indicate some hidden properties of the SM.

\section{Phenomenology of the Standard Model
\label{sect:Pheno}}

Let us discuss input parameters of the SM. It was convenient to count their
number in the \emph{primary} form of the Lagrangian. But for \emph{practical}
applications we use different sets, see \eg \Tref{tab:params}.
Various \emph{EW schemes} with different sets of practical input parameters
are possible (and actually used), since there are relations between them. 
One should keep in mind that the result of calculations does depend on 
the choice because we usually 
work in a limited order of the perturbation theory, while the true relations
between the parameters (and between observed quantities) involve the complete
series. So simple relations appear only at the lowest order, quantum effects 
(radiative corrections) make them complicated.

\begin{table}[h]
\caption{Input parameters of the SM.}
\label{tab:params}
\centering\small
\begin{tabular}{rccccccc|c}
\hline\hline
18(19)=       & 1      & 1   & 1        & 1              & 1       & 9   & 4      & (1)
\\ \hline
primary:      &$g'$    &$g$  & $g_s$    &$m_\Phi$         &$\lambda$&$y_f$&$y_{jk}$ & $\theta_{\mathrm{CP}}$
\\
practical:    &$\alpha$&$M_W$&$\alpha_s$&$G_{\mathrm{Fermi}}$& $M_H$   &$m_f$&$V_{CKM}$& 0 \\
\hline\hline
\end{tabular}
\end{table}

A comprehensive up-to-date set of the SM parameters can be found in the
Review of Particle Physics published by the Particle Data Group 
Collaboration~\cite{Patrignani:2016xqp}.
Let us look at some values of input parameters extracted from experiments:
\begin{itemize}
    \item The \emph{fine structure constant}:
$\alpha^{-1}(0) = 137.035 999 074(44)$ from $(g-2)_e$;
    \item The SM predicts $M_W=M_Z\cos\theta_w\ \ \Rightarrow\ \ M_W<M_Z$, we have now \\
$M_Z = 91.1876(21)$~GeV from LEP1/SLC, 
$M_W = 80.385(15)$~GeV from LEP2/Tevatron/LHC; 
    \item The Fermi coupling constant: 
$G_{\mathrm{Fermi}} = 1.166 378 7(6)\cdot 10^{-5}$~GeV$^{-2}$ from muon decay, 
    \item The top quark mass: $m_t = 173.1(6)$~GeV from Tevatron/LHC;
    \item The Higgs boson mass: $M_H = 125.09(21)(11)$~GeV from {ATLAS \& CMS} (March 2015).
\end{itemize}
One can see that the precision in definition of the parameters varies by several orders of
magnitude. That is related to experimental uncertainties and to the limited accuracy 
of theoretical calculations which are required to extract the parameter values from the 
data. 

\emph{QUESTION: What parameter of the canonical, \ie without neutrino masses and mixing
SM is known now with the least precision?}

\subsection{The muon decay}

\begin{figure}
\centering\includegraphics[width=.25\linewidth]{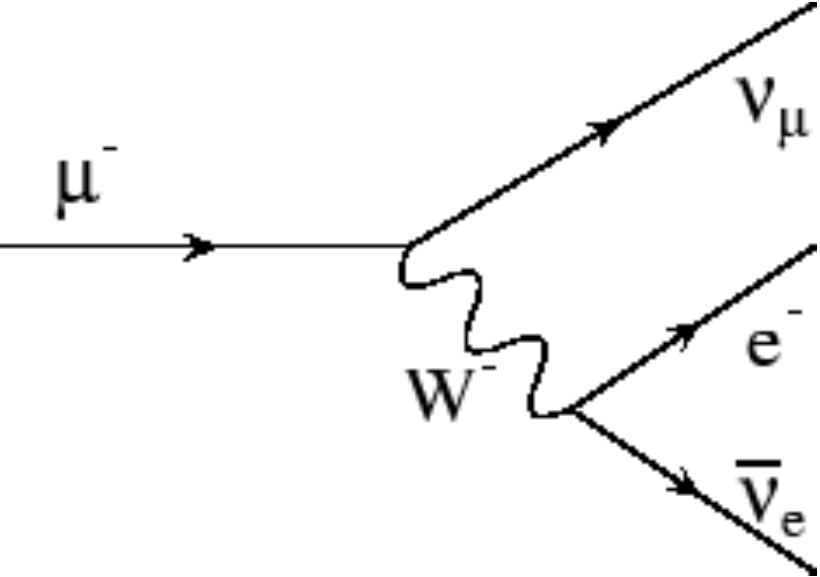}
\caption{The Feynman diagram for muon decay in the SM.}
\label{fig:mu_decay_1}
\end{figure}

Let us consider a few examples of particle interaction processes and start with 
the muon decay $\mu^-\to e^- + \bar{\nu}_e + \nu_\mu$, see~\Fref{fig:mu_decay_1}.  
It is the most clean weak-interaction process. One can say that this process is
one of keystones of particle physics. The muon decay width reads
\begin{eqnarray} \nonumber
&& \Gamma_\mu=\frac{1}{\tau_\mu}=\frac{G_{\mathrm{Fermi}}^2m_\mu^5}{192\pi^3}\biggl[f(m_e^2/m_\mu^2)
+ \mathcal{O}(m_\mu^2/M_W^2) + \mathcal{O}(\alpha)\biggr],
\\ \nonumber
&& f(x) = 1 - 8x + 8x^3 - x^4 - 12x^2\ln x,
\\ \nonumber
&& \mathcal{O}(m_\mu^2/M_W^2) \sim 10^{-6},\qquad 
\mathcal{O}(\alpha) \sim 10^{-3},
\end{eqnarray} 
where $\mathcal{O}(\alpha)$ includes effects of radiative corrections due to loop
(virtual) effects and real photon and/or $e^+e^-$ pair emission.

As mentioned above, the value of the Fermi coupling constant is extracted
from the data on the muon lifetime,
$G_{\mathrm{Fermi}} = 1.166 378 7(6)\cdot 10^{-5}\ \mathrm{GeV}^{-2}$.
The high precision is provided by a large experimental statistics, 
low systematical errors of the final state electron observation,
and by accurate theoretical calculations of radiative corrections.
But impressive precision $(\sim 1\cdot 10^{-6})$ in the measurement 
of the muon life time doesn't give by itself any valuable test of the SM. 
\emph{QUESTION: Why is that so?}
On the other hand, studies of differential distributions in electron 
energy and angle do allow to test the $V-A$ structure of weak interactions 
and look for other possible types of interactions which can be parameterized
in a model-independent way by the so-called \emph{Michel parameters}.

\subsection{Electron and muon anomalous magnetic moments}

The Dirac equations predict gyromagnetic ratio $g_f = 2$ 
in the fermion magnetic moment $\vec{M} = g_f \frac{e}{2m_f}\vec{s}$.
Julian Schwinger in 1948 found that one-loop QED corrections to the vertex function
give the so-called \emph{anomalous magnetic moment}:
\begin{eqnarray}
a_f \equiv \frac{g_f - 2}{2} \approx \frac{\alpha}{2\pi} = 0.001\ 161\ \ldots
\end{eqnarray}
For the electron case, the Harvard experiment~\cite{Hanneke:2010au} obtained
\begin{eqnarray}\nonumber
a^{\mathrm{exp}}_e= 1\ 159\ 652\ 18{\bf 0.73}\ (28) \cdot 10^{-12}
\ \ [0.24\mathrm{ppb}].
\end{eqnarray}
The SM predicts~\cite{Aoyama:2012wj}
\begin{eqnarray}\nonumber
a^{\mathrm{SM}}_e= 1\ 159\ 652\ 18{\bf 1.643}
\ (25)_{8th}(23)_{10th}(16)_{EW+had.}(763)_{\delta\alpha} \cdot 10^{-12}.
\end{eqnarray}
The perfect agreement between the measurement and the theoretical prediction
is a triumph of Quantum electrodynamics. In particular, we note that
$a_f\neq 0$ is a pure quantum loop effect which is absent as in classical physics
as well as in Quantum mechanics.

It is worth to note that the extremely high precision in the experimental
measurement of the electron anomalous magnetic moment allows to use it
as a reference point for definition of the fine structure constant:
$a^{\mathrm{exp}}_e \Rightarrow \alpha^{-1}(0) = 137.035 999 074(44)$.

For the anomalous magnetic moment of muon, the E821 experiment at BNL in 2006 
published the following result of data analysis:
\begin{eqnarray} \nonumber
a^{\mathrm{exp}}_\mu= 116\ 59{\bf 2\ 089}\ (54)(33)\cdot 10^{-11}\ \ \ [0.5\mathrm{ppm}].
\end{eqnarray}
The corresponding theoretical value and the difference are
\begin{eqnarray}
&& a^{\mathrm{SM}}_\mu= 116\ 59{\bf 1\ 840}\ (59)\cdot 10^{-11}\ \ \ [0.5\mathrm{ppm}]
\\ \nonumber
&& \Delta a_\mu \equiv a^{\mathrm{exp}}_\mu - a^{\mathrm{SM}}_\mu = 
{\bf 249}\ (87)\cdot 10^{-11}\ \ \ [\sim 3\sigma].
\end{eqnarray}
First, one can see that both experimental and theoretical values are very accurate.
Second, there is a discrepancy of the order of three standard deviations. That
is a rather rare case for the SM tests. Moreover, this discrepancy remains 
for a long period of time in spite of intensive efforts of experimentalists 
and theoreticians.

The SM prediction consists of the QED, hadronic, and weak contributions: 
\begin{eqnarray} 
&& a_\mu = a_\mu(\mathrm{QED}) + a_\mu(\mathrm{hadronic}) + a_\mu(\mathrm{weak}),
\\ \nonumber
&& a_\mu(\mathrm{QED}) = 116\ 584\ 718\ 845\ (9)(19)(7)(30) \cdot 10^{-14}
\qquad [\mathrm{5\ loops}],
\\ \nonumber
&& a_\mu(\mathrm{hadronic}) =  a_\mu(\mathrm{had.\ vac.pol.}) + a_\mu(\mathrm{had.\ l.b.l}),
=  6949\ (37)(21)\cdot 10^{-11} + 116\ (40)\cdot 10^{-11}, 
\\ \nonumber
&& a_\mu(\mathrm{weak}) = 154\ (2) \cdot 10^{-11} \qquad [\mathrm{2\ loops}].
\end{eqnarray}
Note that the QED contribution to the muon anomalous magnetic moment is essentially the
same as the one to the electron magnetic moment. The only difference is coming from
the dependence on electron and muon masses. As concerning the hadronic and weak interaction
contributions, they are enhanced by the factor $m_\mu^2/m_e^2$ with respect to the electron 
case. The same factor typically appears for hypothetical contributions of new interactions
beyond the SM. For this reason anomalous magnetic moments of muon and tau lepton are
potentially more sensitive to new physics contributions.

One can see that the difference between the theoretical prediction and the experimental
data is almost twice the contribution of weak interactions:
$\Delta a_\mu \sim 2 \times a_\mu(\mathrm{weak})$.
Here by 'weak' we mean the complete electroweak calculation minus the pure QED
contribution. The weak interactions have been directly tested with high precision
experimentally. So it is not so simple to attribute the difference to an
effect of new physics. Nevertheless, there is a bunch of theoretical models
that try to resolve the problem by introduction of new interactions and/or new 
particles.

\subsection{Vacuum polarization}

\begin{figure}
\centering\includegraphics[width=.25\linewidth]{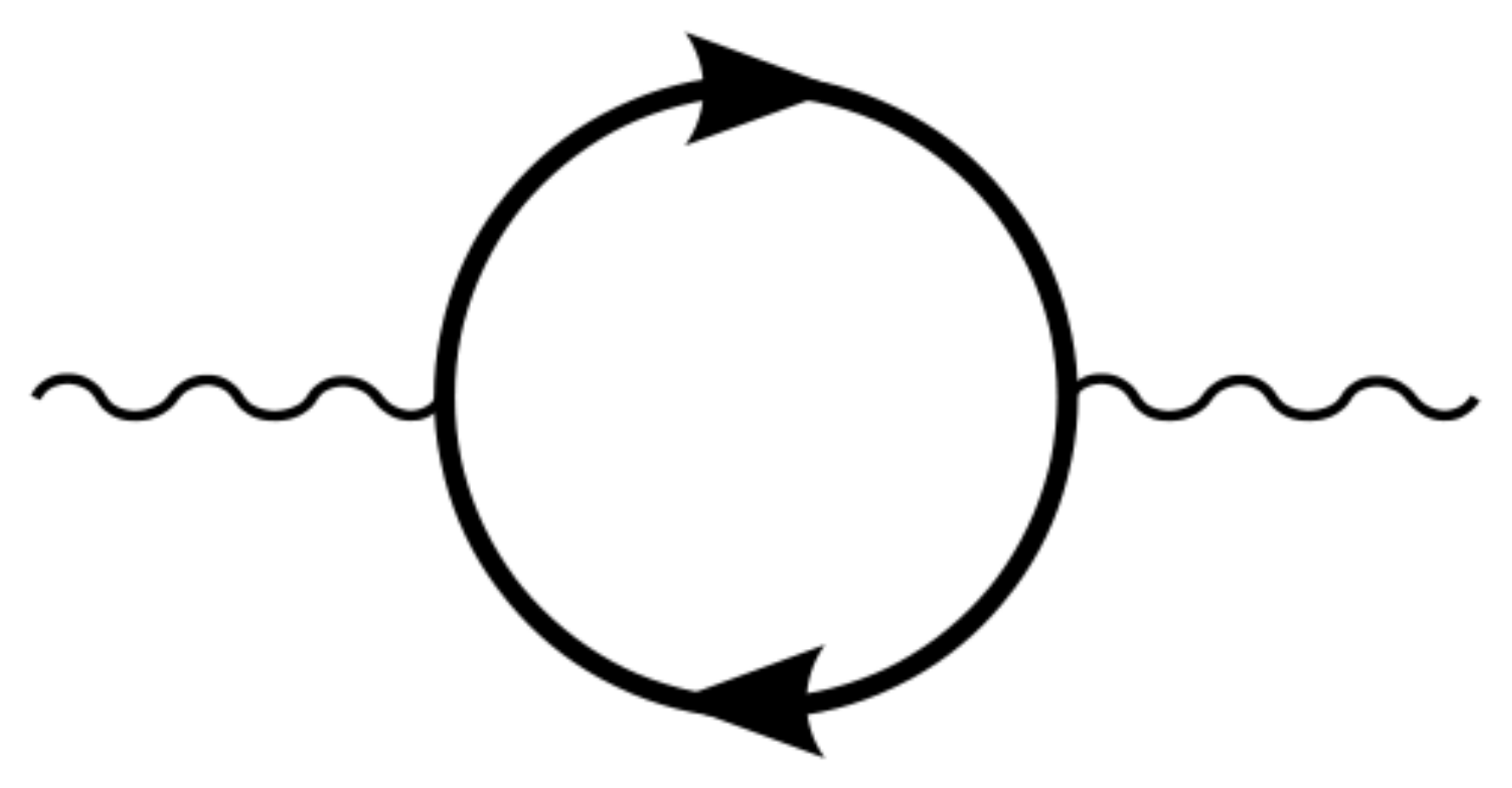}
\caption{The one-loop Feynman diagram for QED vacuum polarization.}
\label{fig:vac_pol}
\end{figure}

By direct calculation in QED, one can see that virtual charged fermion 
anti-fermion pairs provide a screening effect for the electric force 
between probe charges. 
Resummation of bubbles, see~\Fref{fig:vac_pol}, gives 
\begin{eqnarray} \nonumber
&& \alpha(q^2) = \frac{\alpha(0)}{1-\Pi(q^2)}, \qquad 
\mathrm{e.g.}\ \   \alpha^{-1}(M_Z^2) \approx 128.944(19),
\\ \nonumber
&& \Pi(q^2) = \frac{\alpha(0)}{\pi}\biggl(
\frac{1}{3}\ln\left(\frac{-q^2}{m_e^2}\right) 
- \frac{5}{9} + \delta(q^2) \biggr)
+ \mathcal{O}(\alpha^2),
\\ 
&& \delta(q^2) = \delta_{\mu}(q^2) + \delta_{\tau}(q^2) + \delta_{W}(q^2)
+ \delta_{\mathrm{hadr.}}(q^2).
\end{eqnarray}
The hadronic contribution to vacuum polarization $\delta_{\mathrm{hadr.}}(q^2)$ 
for $|q^2|$ below a few GeV$^2$
is not calculable within the perturbation theory. Now we get it
from experimental data on $e^+e^-\to\mathrm{hadrons}$ and 
$\tau\to\nu_\tau+\mathrm{hadrons}$ with the help of dispersion relations,
see \eg review~\cite{Actis:2010gg}. 
Lattice results for this quantity are approaching.

Note that screening, \ie an effective reduction of the observed
charge with increasing of distance, is related to the minus
sign attributed to a fermion loop in the Feynman rules.

\emph{ QUESTION: Estimate the value of $q^2_0$ at which $\alpha(q^2_0)=\infty$.}

This singularity is known as the \emph{Landau pole}. Formally, such a behaviour
of QED brakes unitarity at large energies. But that happens at energies much
higher than any practical energy scale including the Planck mass and the mass
of the visible part of the Universe. So we keep this problem in mind as a theoretical
issue which stimulates our searches for a more fundamental description of Nature.

\subsection{Experimental tests of the SM at LEP}

\begin{figure}
\centering\includegraphics[width=.5\linewidth]{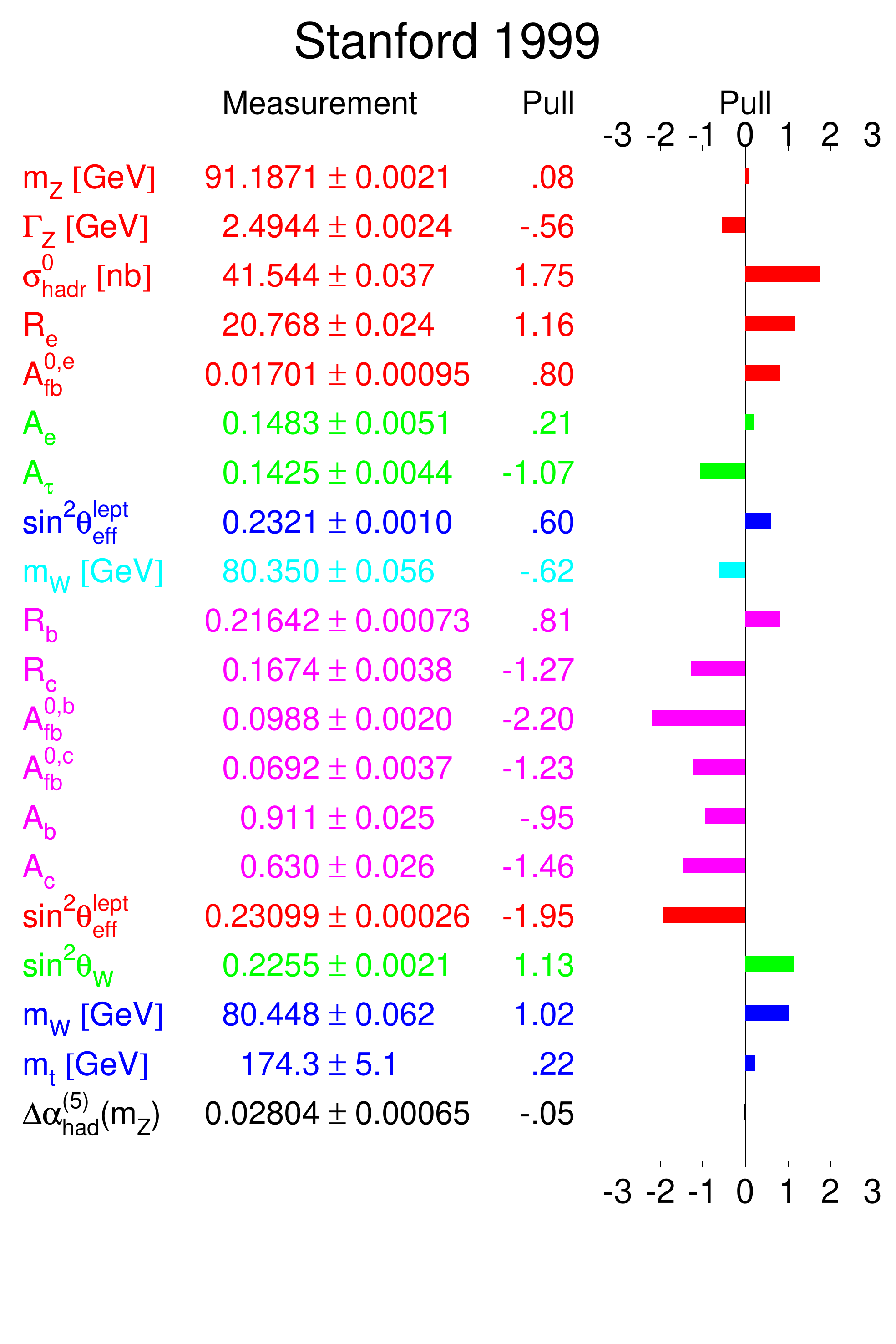}
\caption{Pulls of pseudo-observables at LEP~\cite{LEPEWWG}.}
\label{fig:pulls}
\end{figure}

After the analysis of LEP1 and LEP2 experimental data, the LEP Electroweak Working Group 
(LEPEWWG)~\cite{LEPEWWG} illustrated 
the overall status of the Standard Model by the so-called \emph{pulls}, 
see~\Fref{fig:pulls}.
The pulls are defined as differences between the measurement and 
the SM prediction calculated for the central values of the fitted SM input parameters
$[\alpha(M_Z^2)=1/128.878$,\ $\alpha_s(M_Z^2)=0.1194$,\ $M_Z=91.1865\,\mbox{GeV}$,\
$m_t=171.1\,\mbox{GeV}]$ divided by the experimental error.
Although there are several points where deviations between the theory 
and experiment approach two standard deviations,
the average situation should be ranked as extremely good. 
We note that the level of precision reached is of the order of $\sim 10^{-3}$,
and that it is extremely non-trivial to control all experimental systematics
at this level. 

\begin{figure}
\centering\includegraphics[width=.8\linewidth]{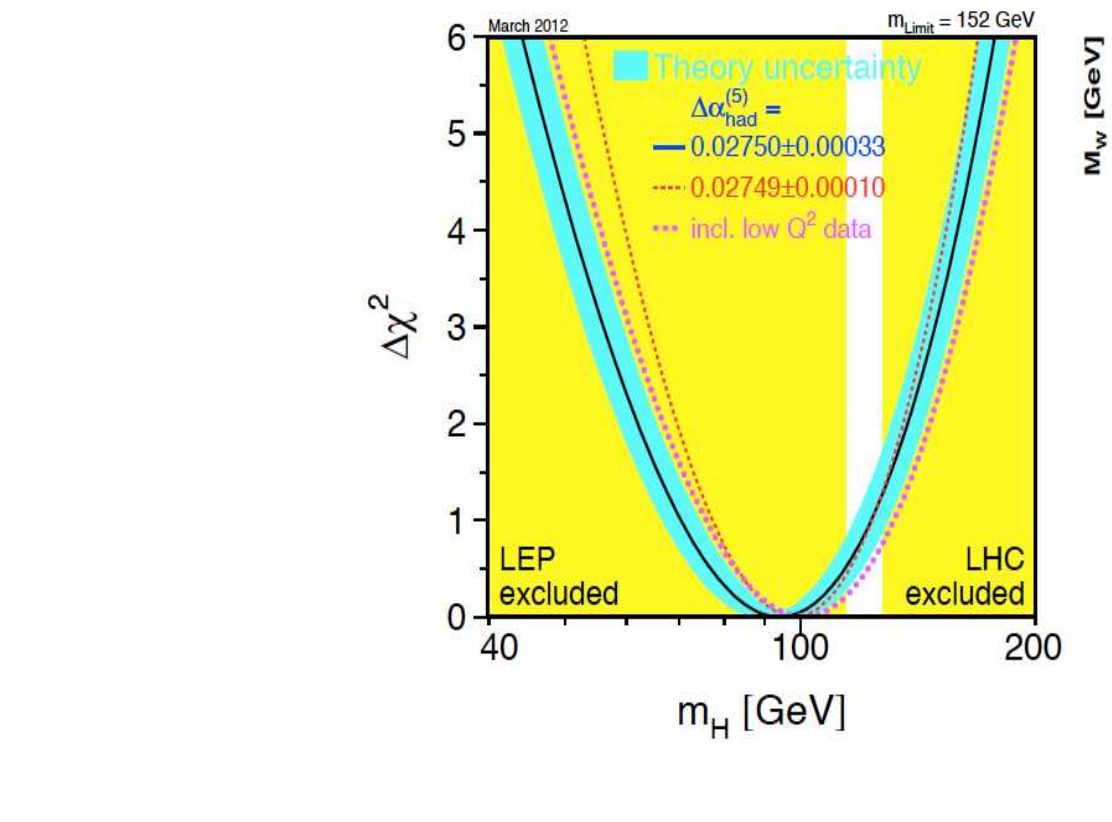}
\caption{The curve shows $\Delta\chi^2_{\rm{min}}(M_H^2)=\chi^2_{\rm{min}}(M_H^2)-\chi^2_{\rm{min}}$
as a function of $M_H$. 
The width of the shaded band around the curve shows the theoretical uncertainty. 
The vertical bands show the 95\% CL exclusion limit on $M_H$ from the direct searches
at LEP (left) and at LHC (right). The dashed curve is the result
obtained using the evaluation of $\Delta\alpha^{(5)}(M_Z^2)$. The dotted curve corresponds to a
fit including also the low-$Q^2$ data.}
\label{fig:blue_band}
\end{figure}

Through quantum effects the observed cross sections of electron-positron
annihilation at LEP depend on all parameters of the Standard Model including
the Higgs boson mass. The so-called yellow band plot~\Fref{fig:blue_band} 
shows the fit of $M_H$ performed by LEPEWWG~\cite{LEPEWWG} 
with the LEP data in March 2012.
The left yellow area has been excluded by direct 
searches at LEP, and the right one was also excluded by LHC. 
The plot is derived from a combined fit of all the world experimental data to 
the SM exploiting the best knowledge of precision theoretical calculations which 
is realized in computer codes {\tt ZFITTER}~\cite{Arbuzov:2005ma} 
and {\tt TOPAZ0}~\cite{Montagna:1998kp}.
One can see that the data was not very sensitive to $M_H$, but the fit
unambiguously prefers a relatively light Higgs boson. Now we can say that the
measured value of this parameter agrees very well with the LEP fit. That indirectly
confirms again the consistency and the power of the Standard Model.

\begin{figure}
\centering\includegraphics[width=.8\linewidth]{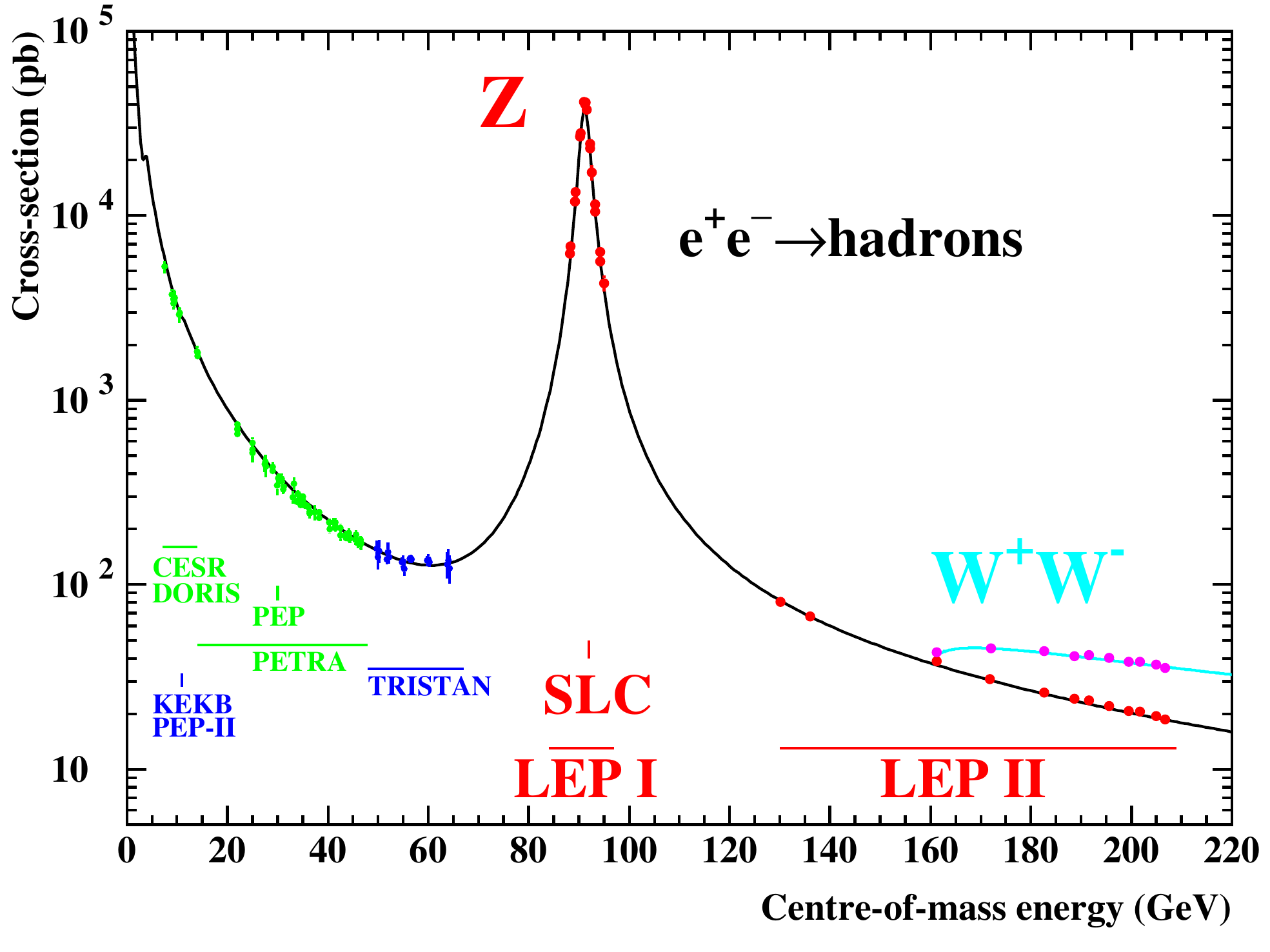}
\caption{Measurements of the $e^+e^-\to\mathrm{hadrons}$\  cross section.}
\label{fig:hadr_LEP}
\end{figure}

It is interesting also to look at the behavior of the cross sections of electron-positron
annihilation into hadrons as a function of energy~\Fref{fig:hadr_LEP}. A clear peak
at the $Z$ boson mass is seen. The excellent agreement of the experimental data
with the SM predictions is achieved only after inclusion of QCD and electroweak
radiative corrections which reach dozens of percent in the vicinity of the peak.

\begin{figure}
\centering\includegraphics[width=.8\linewidth]{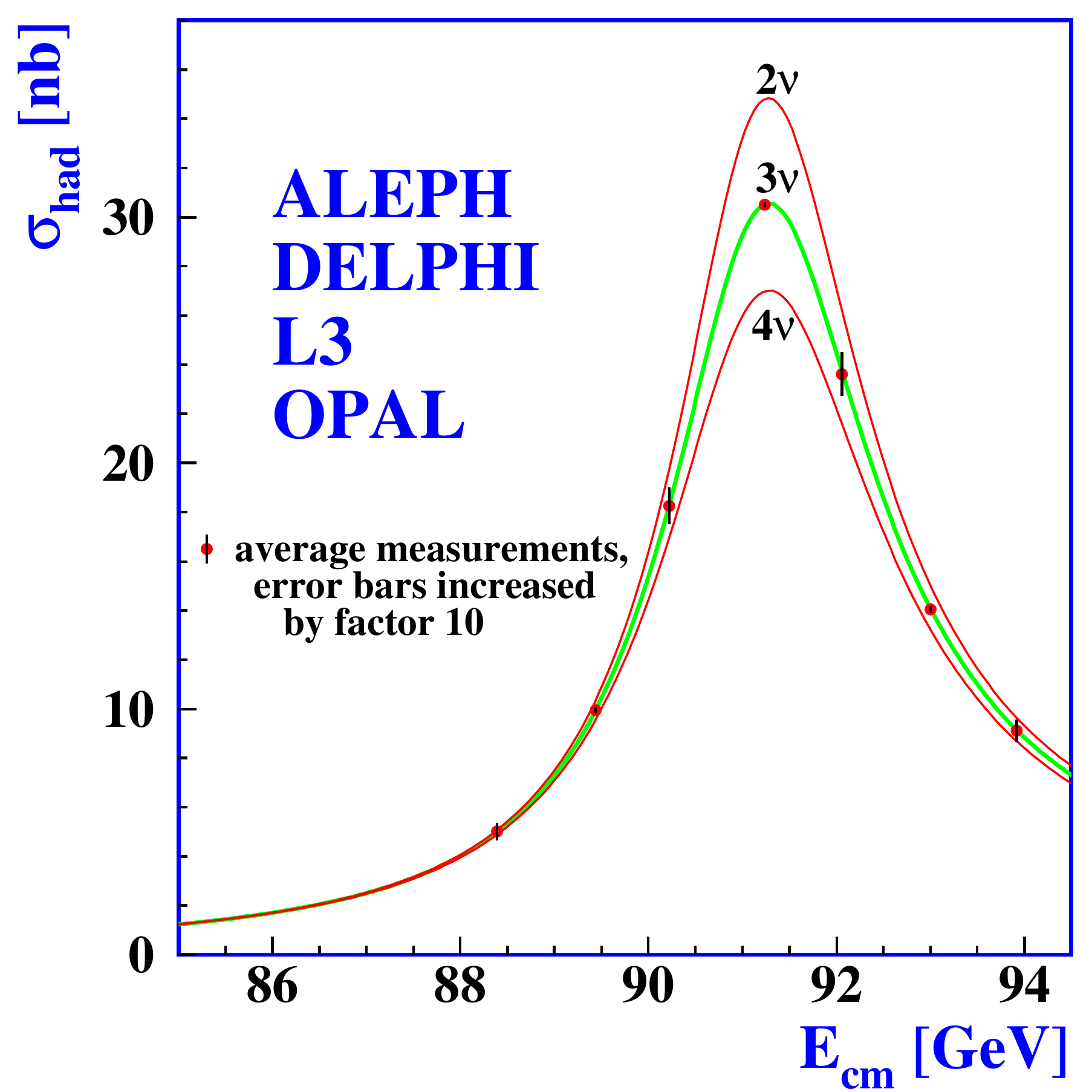}
\caption{Annihilation cross section $e^+e^-\to\mathrm{hadrons}$ for different numbers 
of light neutrinos: measured distribution vs. the SM prediction.}
\label{fig:nu_LEP}
\end{figure}

A peculiar result was obtained at LEP for the number of (light) neutrinos, 
see~\Fref{fig:nu_LEP}.
Even so that the final state neutrinos in the process $e^+ + e^- \to Z \to \nu+\bar{\nu}$
was not observed, the corresponding cross section was restored with the help
of the separately measured hadronic and leptonic cross sections~\cite{LEPEWWG},
and the total $Z$ boson width.

\begin{figure}
\centering\includegraphics[width=.8\linewidth]{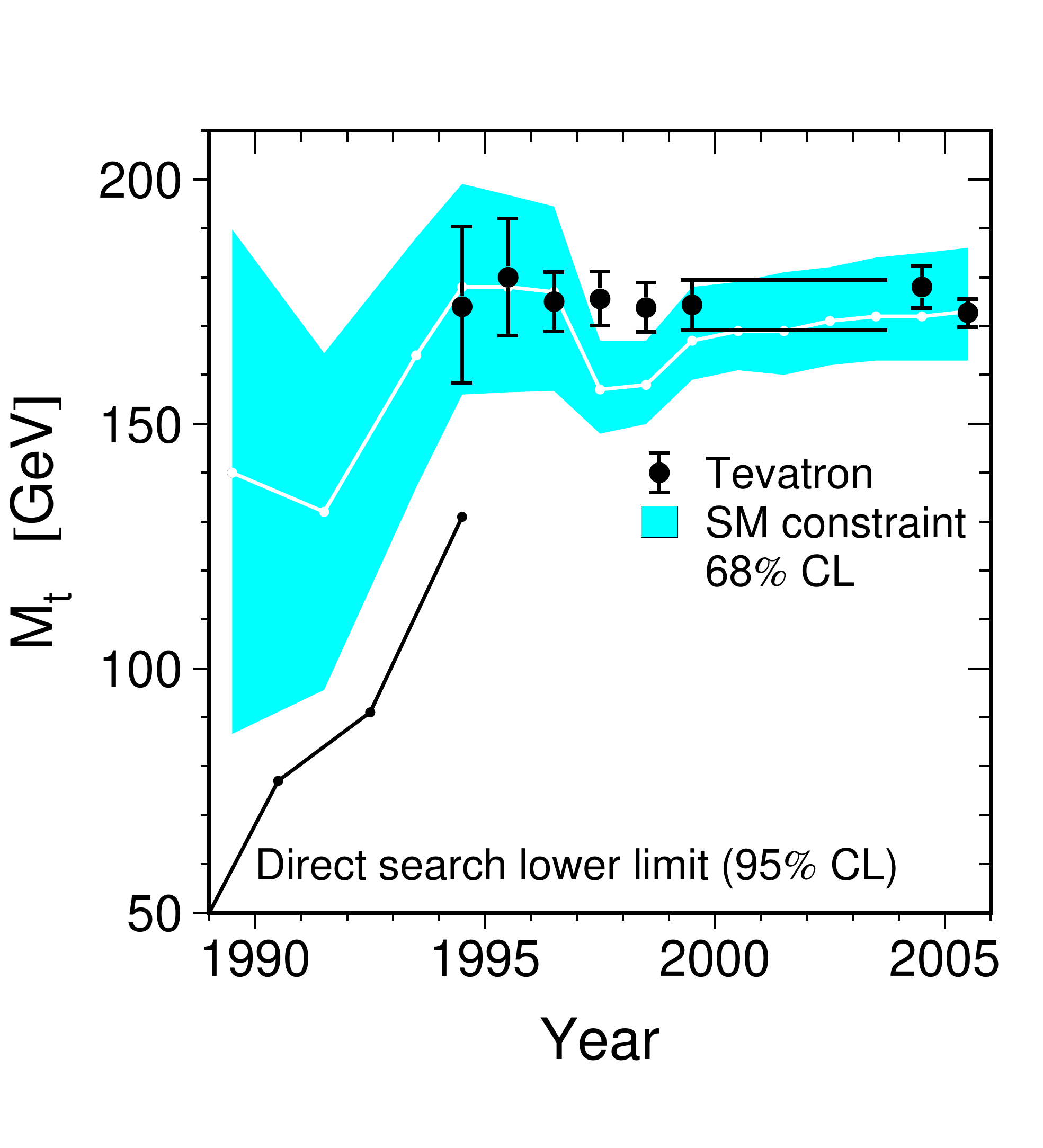}
\caption{Indirect (LEP) and direct (Tevatron) measurements of the top quark mass.}
\label{fig:top_LEP}
\end{figure}

It appears that the dependence of LEP observables on quantum loop effects 
involving top quark is rather strong. So even without approaching the
direct production of top quark, LEP experiments were able to extract
information about its mass. The top quark mass 'history' (till 2006) 
is shown by~\Fref{fig:top_LEP}.

\begin{figure}
\centering\includegraphics[width=.4\linewidth]{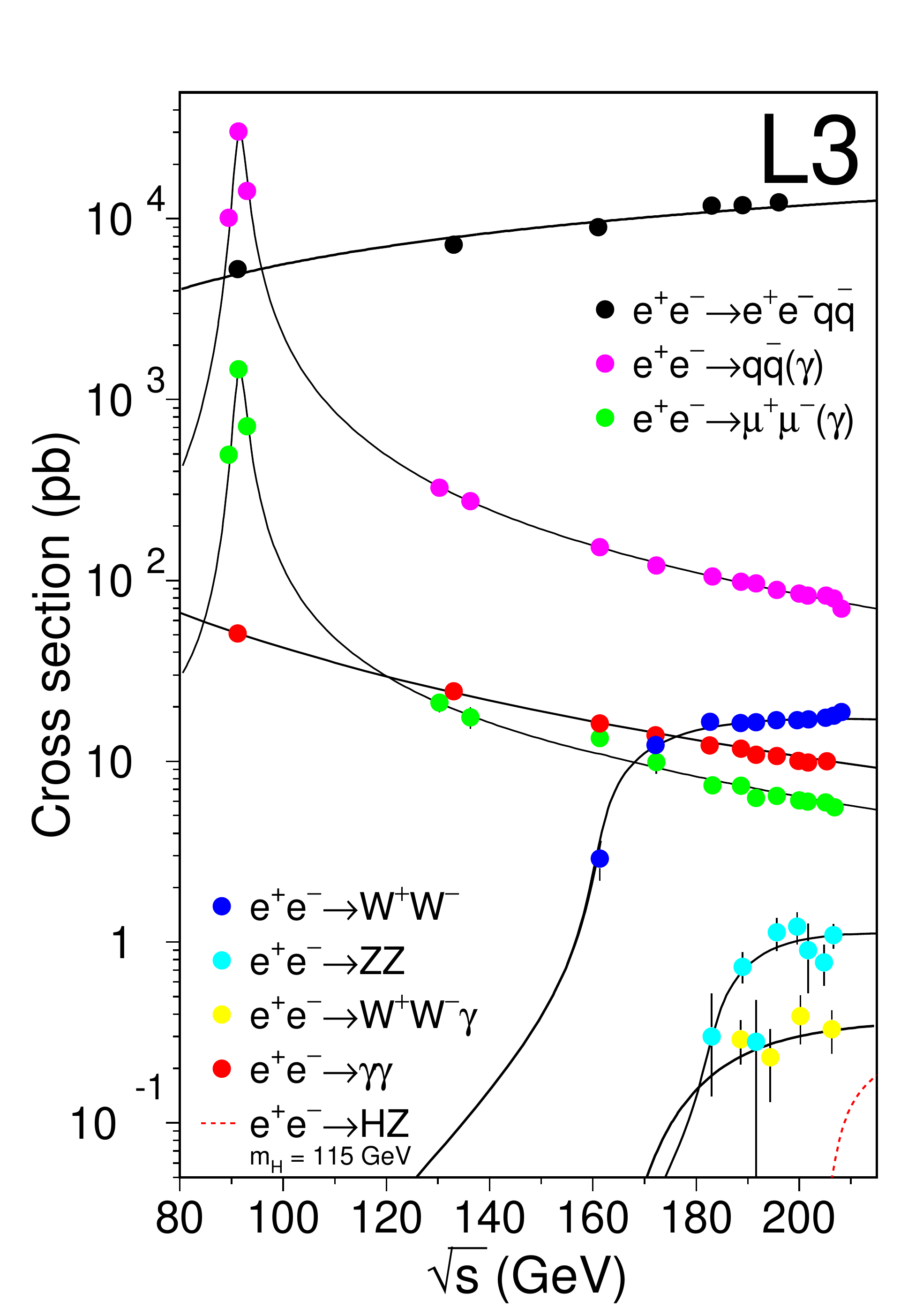}
\caption{Cross-sections of electroweak SM processes at LEP2.}
\label{fig:LEP2}
\end{figure}

In general, all LEP measurements of various cross-sections of electroweak 
SM processes were found in a very good agreement with theoretical predictions
obtained within the SM, see plot~\Fref{fig:LEP2} from the LEPEWWG~\cite{LEPEWWG} 
2013 report. 
The dots show the measurements and curves are the SM predictions with radiative
corrections taken into account.

\subsection{Measurements of SM processes at LHC}

\begin{figure}
\centering\includegraphics[width=.8\linewidth]{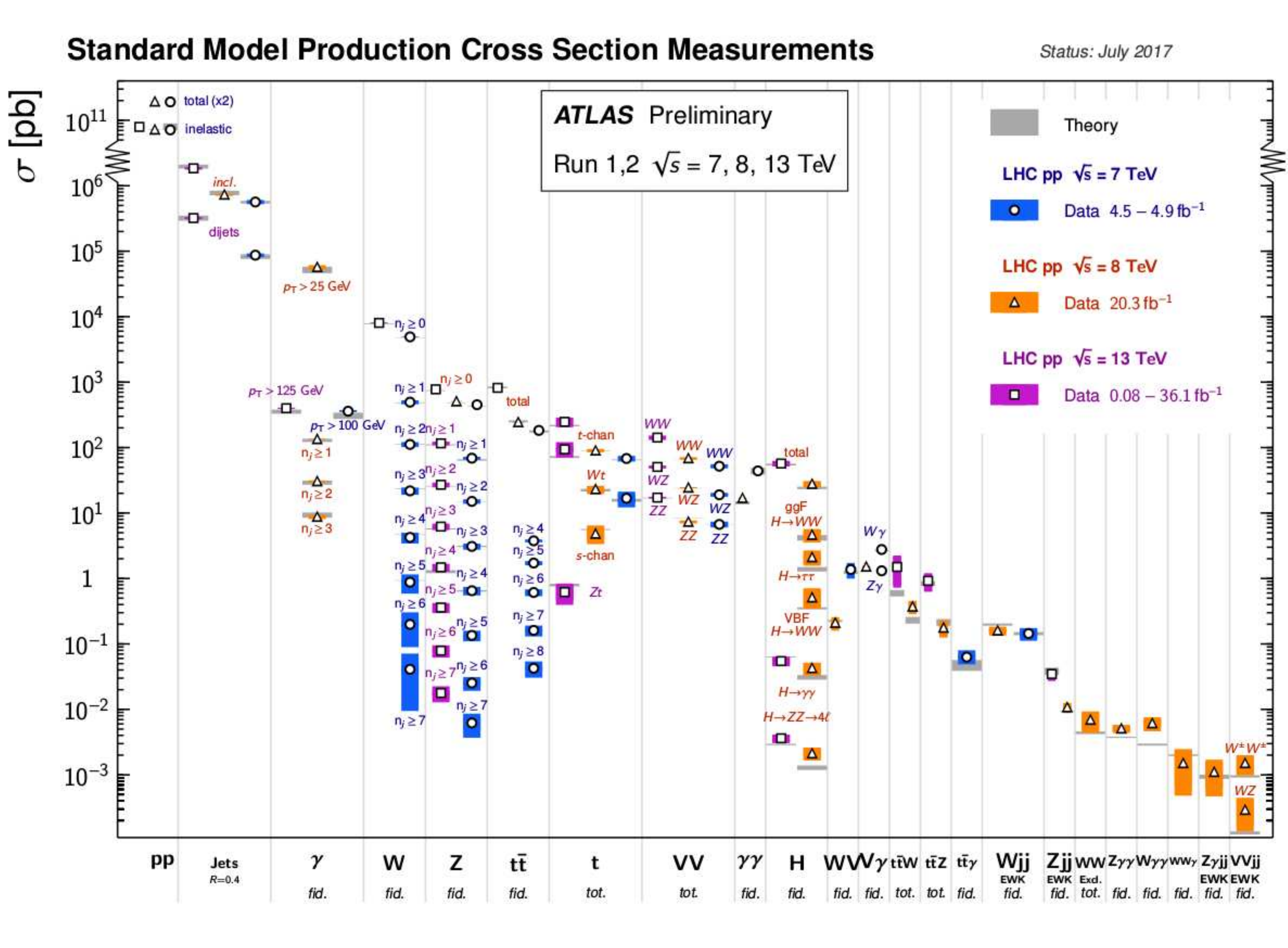}
\caption{SM cross sections measured by ATLAS (public results).}
\label{fig:ATLAS}
\end{figure}

\begin{figure}
\centering\includegraphics[width=.8\linewidth]{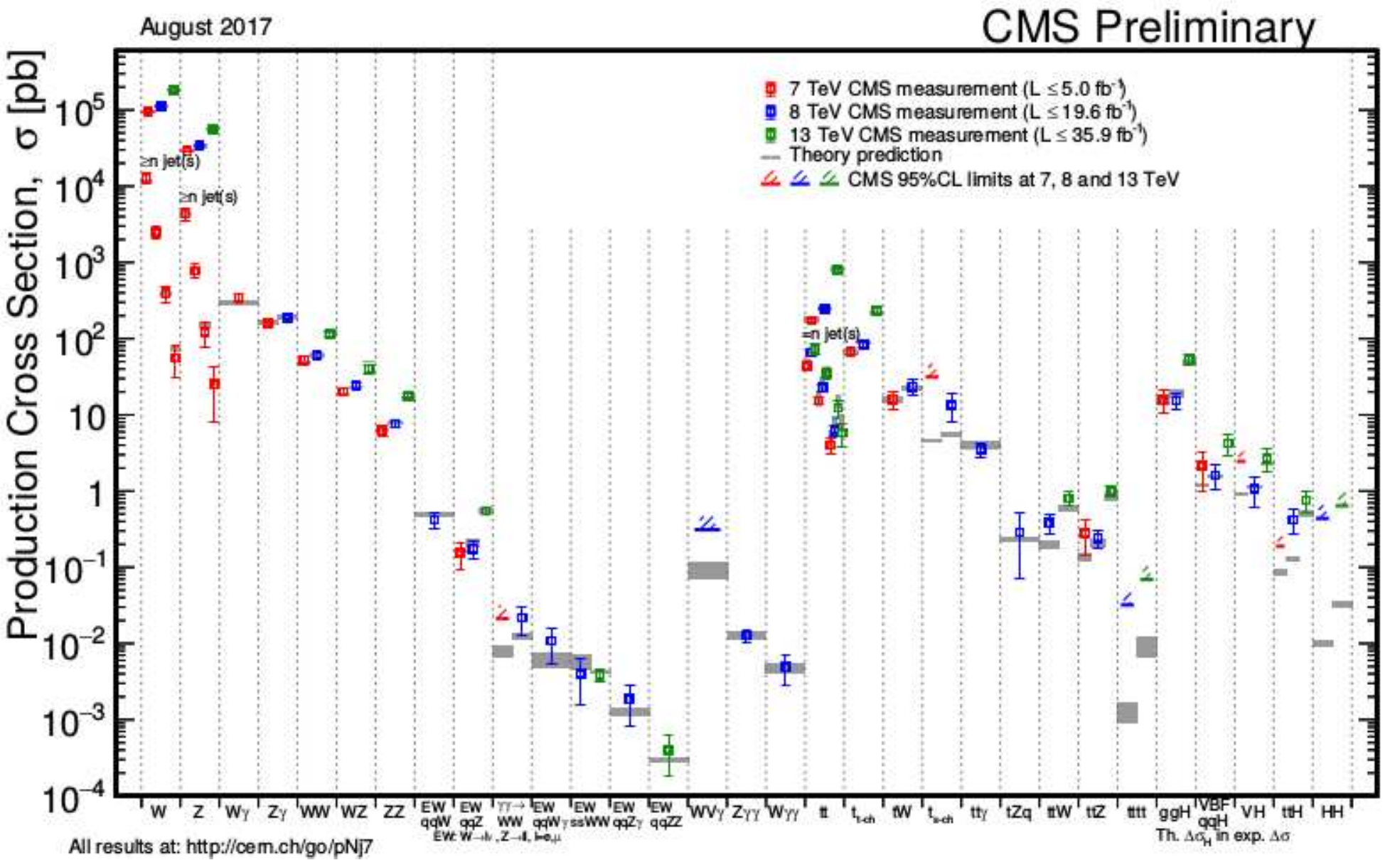}
\caption{SM cross sections measured by CMS (public results).}
\label{fig:CMS}
\end{figure}

The Large Hadron Collider at CERN is not only a discovery machine. In fact the
large luminosity and advanced detectors allow to perform there high-precision
tests of the Standard Model. High statistics on many SM processes is collected.
Plots \Fref{fig:ATLAS} and \Fref{fig:CMS} show the public preliminary results
of the ATLAS and CMS collaborations. One can see that we have again a good
agreement for all channels. Certainly, the tests of the SM will be continued
at LHC at higher energies and luminosity. That is one of the main tasks the
LHC physical programme. The proton-antiproton collider Tevatron has proven that hadronic
colliders can do high-precision studies of the SM. In particular, CDF and D0 experiments
at Tevatron managed to exceed LEP in the precision of the $W$ boson mass measurement.

\begin{figure}
\centering\includegraphics[width=.5\linewidth]{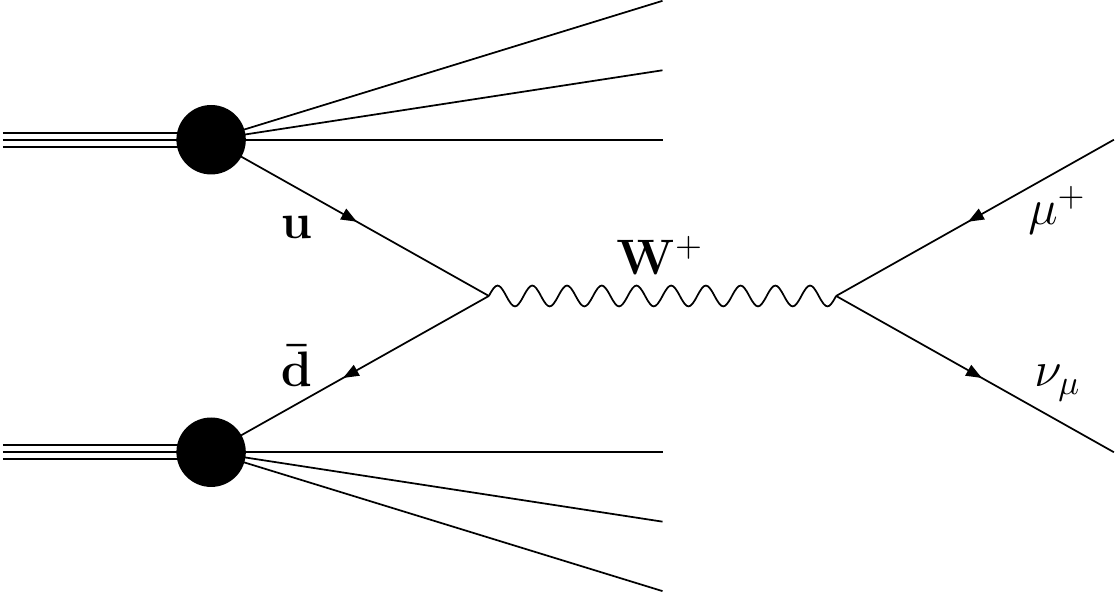}
\caption{Schematic Feynman diagram for the charged current Drell--Yan-like process.}
\label{fig:DY}
\end{figure}

At LHC the best precision in SM processes measurement is reached for
the Drell--Yan-like processes. A schematic diagram for such a process is shown 
on~\Fref{fig:DY}. These processes are distinguished by production of final state
leptons which can be accurately detected. We differ the neutral current (NC)
Drell--Yan-like processes which involve intermediate $Z$ bosons and photons,
and the charged current (CC) ones which go through $W^\pm$ bosons. The main contribution
to the (observed) total cross section of these processes comes from the domain
where the invariant mass of the final state lepton pair is close to the masses
of $Z$ and $W$ bosons. So these processes are also known as single $Z$ and $W$
production reactions.
The CC and NC Drell--Yan-like processes at LHC are used for:
\begin{itemize}
    \item luminosity monitoring;
    \item $W$ mass and width measurements;
    \item extraction of parton density functions;
    \item detector calibration;
    \item background to many other processes;
    \item and new physics searches. 
\end{itemize}
In particular, a new peak in the observed invariant-mass distribution of final leptons
can indicate the presence of a new intermediate particle.

\section{Conclusions
\label{sect:Conclusions}}

Let us summarize the status of the SM. We see that it is a rather elegant
construction which allows making systematic predictions for an extremely wide 
range of observables in particle physics. The energy range of its applicability
covers the whole domain which is explored experimentally while 
the true limits remain unknown. We do not understand all features of the model,
the origin(s) of its symmetries and parameter values. But we see that the SM
has the highest predictive power among all models in particle physics
and it successfully passed verification at thousands of experiments.

There are several particularly nice features of the SM:
\begin{itemize}
    \item it is renormalizable and unitary $\Rightarrow$ it gives finite predictions;
    \item its predictions do agree with experimental data;
    \item symmetry principles are extensively exploited;
    \item it is minimal;
    \item all its particles are discovered;
    \item the structure of interactions is fixed (but not yet tested everywhere);
    \item not so many free parameters, all are fixed;
    \item CP violation is allowed;
    \item tree-level flavor-changing neutral currents are not present;
    \item there is a room to incorporate neutrino masses and mixing.
    \end{itemize}
In principle in the future, the SM can be embedded into a more general theory 
as an effective low-energy approximation. But in any case the SM will remain
the working tool in the energy domain relevant for the absolute majority
of our experiments.

For many reasons we do not believe that the SM is the final 
'theory of everything'. Of course first of all, we have to mention that
the SM is not joined with General Relativity. But frankly speaking, that is  
mostly the problem of GR, since the SM itself is ready to be incorporated into
a generalized joint QFT construction, while GR is not (yet) quantized. 
The naturalness problem discussed above in Sect.~\ref{sect:natural} 
indicates that either some new physics should be 
very close to the EW energy scale, or we do not understand features
of the renormalization procedure in the SM. 
In general, we have a lot of open questions within the SM:
\begin{itemize}
    \item the origin of symmetries;
    \item the origin of EW and QCD energy scales;
    \item the origin of 3 fermion generations;
    \item the origin of neutrino masses;
    \item the hierarchy of lepton masses;
    \item the absence of strong CP violation in the QCD sector;
    \item confinement in QCD, and so on\ldots
\end{itemize}
There are also some phenomenological issues:
\begin{itemize}
    \item the baryon asymmetry in the Universe;
    \item the dark matter;
    \item the dark energy;
    \item the proton charge radius, $(g-2)_\mu$, and not much else\ldots
\end{itemize}
The first three items above are related to Cosmology, 
see the corresponding lecture course. We should note also that
most of observations in Cosmology and Astrophysics are well described
within the Standard Model (and General Relativity). But for the listed cases
we need most likely something beyond the SM. The last item in the list 
claims that there are some tensions in the predictions of the SM and 
measurements at experiments in particle physics. 

So we see that the SM is build using some nice fundamental principles but 
also with a substantial phenomenological input.
The most \emph{valuable task} for high-energy physicists now is to find the 
limits of the SM applicability energy domain.
Yes, we hope to discover soon new physical phenomena. But any kind of new physics
ought to preserve the correspondence to the SM.
The SM contains good mechanisms to generate masses of vector bosons
and fermions, but it doesn't show the \emph{origin(s)} of the 
electroweak and QCD energy scales. 

So, the SM can not be the full story in particle physics, we still have 
a lot to explore. Good luck!

%\subsection*{Acknowledgements}
%
%If required, acknowledgements should appear as an unnumbered
%subsection immediately before the references section.

\end{document}